\documentclass[twocolumn]{aastex62}
\usepackage{natbib}
\usepackage{amsmath}

\newcommand{\angstrom}{\textup{\angstrom}}

\newcommand\Ms{M_{\odot}}

\newcommand\stf{star formation }
\newcommand\sfing {star-forming }

\newcommand\gc{globular cluster }

\newcommand\gcs{globular clusters }

\newcommand\sfe{star formation efficiency }

\newcommand\hhb{hot hydrogen-burning }

\received{2023 Dec 18}
\revised{2024 Jan 25}
\accepted{2024 Feb 1}

\shorttitle{Globular Clusters in Mass-$F_{1P}$ Space}
\shortauthors{Parmentier}

\begin{document}

\title{Cracking the relation between mass and 1P-star fraction of globular clusters: \\
I. Present-day cluster masses as a first tool} 

\correspondingauthor{Genevi\`eve Parmentier}
\author[0000-0002-2152-4298]{Genevi\`eve Parmentier}
\email{gparm@ari.uni-heidelberg.de}
\affiliation{Astronomisches Rechen-Institut, Zentrum f\"ur Astronomie der Universit\"at Heidelberg, M\"onchhofstr. 12-14, D-69120 Heidelberg, Germany}



\begin{abstract}
The phenomenon of multiple stellar populations is exacerbated in massive globular clusters, with the fraction of first-population (1P) stars a decreasing function of the cluster present-day mass.  We decipher this relation in far greater detail than has been done so far. 
We assume (i) a fixed stellar mass threshold for the formation of second-population (2P) stars, (ii) a power-law scaling $F_{1P} \propto m_{ecl}^{-1}$ between the mass $m_{ecl}$ of newly-formed clusters and their 1P-star fraction $F_{1P}$, and (iii) a constant $F_{1P}$ over time.  The $F_{1P}(m_{ecl})$ relation is then evolved up to an age of 12\,Gyr for tidal field strengths representative of the entire Galactic halo.  The 12\,Gyr-old model tracks cover extremely well the present-day distribution of Galactic globular clusters in (mass,$F_{1P}$) space.  The distribution is curtailed on its top-right side by the scarcity of clusters at large Galactocentric distances, and on its bottom-left side by the initial scarcity of very high-mass clusters, and dynamical friction.  Given their distinct dissolution rates, "inner" and "outer" model clusters are offset from each other, as observed.  The locus of Magellanic Clouds clusters in (mass,$F_{1P}$) space is as expected for intermediate-age clusters evolving in a gentle tidal field.  Given the assumed constancy of $F_{1P}$, we conclude that 2P-stars do not necessarily form centrally-concentrated.  We infer a minimum mass of $4 \cdot 10^5\,\Ms$ for multiple-populations clusters at secular evolution onset.  This high-mass threshold severely limits the amount of 2P-stars lost from evolving clusters, thereby fitting the low 2P-star fraction of the Galactic halo field.       
\end{abstract}


\section{Introduction} \label{sec:intro}

Globular clusters are not the simple stellar populations we used to think they are.  The vast majority of them consist of two main populations that bear distinct chemical abundances in light elements \citep[e.g.][]{kra94,car10a,gra12,cha16,bala18}.  Stars of the first population are oxygen- and carbon-rich, nitrogen- and sodium-poor, as the vast majority of halo field stars are.  In contrast, stars of the second population are enhanced in nitrogen and sodium, but depleted in oxygen and carbon.  Light-element abundance variations among \gc stars are therefore not random: they present well-defined correlations (Na-N-He) and anti-correlations (Na-O, N-C), which point to second-population stars having formed out of gas polluted by hot hydrogen-burning products \citep[i.e.~hydrogen burning through CNO-, NeNa-, and MgAl-cycles;][]{lan93}.  Evidence for these abundance variations have been provided by both spectroscopy \citep[see e.g. ][]{car09,smo11,muc12,lar13} and photometry \citep[see e.g.][]{pio07,mil08,and09,mil15}.    

At which stage of their early evolution are \gcs polluted?  What is (are) the polluter(s)?  These questions have so far remained unanswered.  Various stellar polluters -- asymptotic giant branch stars \citep[mass $\simeq 4-8\,\Ms$,][]{dant02,derc08,dant16}, interacting massive binaries \citep[masses $15\,\Ms$-$20\,\Ms$,][]{deM09}, fast-rotating massive stars \citep[mass $> 20\,\Ms$, ][]{pra06,dec07}, Very Massive Stars \citep[mass $10^2-10^3\,\Ms$,][]{vin18,hig23}, and Super Massive Stars \citep[mass $\gtrsim 10^4\,\Ms$, ][]{den14} -- have been proposed, all of them massive enough to experience hot hydrogen burning.  But no consensus has been reached to date \citep{ren15,bala18}.

In Galactic \gcs and Magellanic Clouds clusters, extensive mappings of both populations have been considerably accelerated by the so-called Chromosome Map, a photometric tool that relies on the sensitivity of stellar ultraviolet colors to light-element abundances \citep{mil15}.  With star counts per cluster ranging from several hundreds to a few thousands \citep[see e.g. Table~2 in][]{mil17}, the fraction of stars in either population can be reliably inferred, and trends with individual and environmental cluster properties identified.  

In what follows, we refer to stars of the first population as pristine or 1P-stars, and to stars of the second population as polluted or 2P-stars.    

The multiple-populations phenomenon is exacerbated in more massive globular clusters, with the fraction of pristine stars, $F_{1P}$, a decreasing function of the host-cluster present-day mass.  Equivalently, the fraction of polluted stars increases with the host-cluster mass \citep{mil20}.  Spectroscopic analyses have also shown that the most polluted stars achieve more extreme light-element abundances in more massive clusters \citep{mil17,mil18,zen19,lag19}.  In contrast, Galactic \gcs made of 1P-stars only are all low-mass clusters (present-day mass $\lesssim 3 \cdot 10^4\,\Ms$).    

Although the cluster mass is not the only driver of multiple stellar populations \citep[see e.g.][for the influence of metallicity]{car10a,pan17}, these observations reinforce early and current suggestions that a stellar system must achieve a minimum mass to host multiple populations \citep{car10a,bek11,mil20}.  More specifically, forming clusters must be massive {\it and} compact enough so as to retain their hot hydrogen-burning products \citep{kra16, mba21}.  As pointed out by \citet{sal19}, this probably explains why the field of dwarf galaxies seems devoid of 2P-stars \citep{nor17,sal19}, even though their \gcs host multiple populations (see e.g. \citet{muc09} for the Large Magellanic Cloud, \citet{nie17} for the Small Magellanic Cloud, and \citet{lar14} for the Fornax dwarf spheroidal galaxy). 

It is tempting to fit a straightline on the anti-correlation between the logarithmic present-day mass of \gcs and their fraction of 1P-stars.  Yet, how much of the underlying physics does a linear fit actually capture?  It leaves off Galactic single-population clusters, and most Magellanic Clouds clusters, including multiple-populations ones \citep[see top-panel of fig.~7 in][]{mil20}.  A linear fit also ignores the widening of the distribution towards higher pristine-star fractions.  In fact, distinct linear fits are needed to cover Galactic \gcs with small and large Galactocentric distances, as these are shifted with respect to each other \citep{zen19,mil20}.  

Interestingly, the data seem to show an upturn in the $F_{1P}$-vs.-present-day mass space once Magellanic Clouds clusters and single-population clusters are taken into account \citep[see top-left panel in fig.~7 of][]{mil20}.  The trend is even neater when considering initial cluster mass estimates (see top-right panel in fig.~7 of \citealt{mil20} and middle panel in fig.~11 of \citealt{don21}).  
Guided by this upturn, we assume that the mass and 1P-star fraction of \gcs massive enough to form 2P-stars obey a power-law relation $F_{1P} \propto m_{ecl}^{-1}$ by the end of their formation.  Here, $m_{ecl}$ is therefore the stellar mass of \gcs at residual \sfing gas expulsion.  We shall map how this relation evolves as a function of time and Galactocentric distance, assuming that the 1P-fraction of each cluster remains constant over time.  That is, clusters lose their 1P- and 2P-stars with the same likelihood, and one population does not form more centrally concentrated than the other.  As we shall see, our model already provides a fair account of how \gcs are distributed in the space of their present-day mass versus pristine-star fraction.  

Our approach contrasts with earlier models, in which 2P-stars form more centrally concentrated than their 1P-counterparts, thereby resulting in the preferential tidal stripping of 1P-stars \citep[e.g.][]{dec07,derc08}.  This is a requisite when 2P-stars are assumed to form out of 1P-stars ejecta.  Given that 2P-stars represent today 20-80\,\% of their host cluster mass, 1P-populations must be assumed much more massive than they are today, so as to provide enough ejecta out of which 2P-stars form (aka "mass budget problem").  But the preferential tidal stripping of 1P-populations is then required so as to decrease their mass down to their current value, while preserving the centrally-located 2P-populations \citep{ves10}.  

Interacting Massive Binaries and Super Massive Stars have been suggested as a way around the mass-budget problem , especially when their ejecta are diluted with pristine gas (see \citet{deM09} and \citet{gie18}, respectively, whose scenarios also foresee that 2P-stars form centrally concentrated).  The dilution of 1P-stars ejecta with pristine gas (either in-situ or accreted) is indeed required by the observations of Li in \gc stars \citep{muc11}, including 2P-stars \citep[see][for a discussion]{bala18}.  In this contribution, we remain agnostic regarding the source of the cluster pollution, and we assume that enough polluted gas is provided at a time when the residual \sfing gas still pervades cluster-forming clumps.  

The observed spatial distributions of 1P- and 2P-stars inside \gcs have been much debated.  In some clusters, the 2P-population is more spatially concentrated, as expected from early formation scenarios.  But in some other clusters, this is the 1P-population that is more centrally concentrated (e.g. M15, \citealt{lar15}; NGC 362 and NGC 6723, \citealt{lim16}), which challenges the hypothesis that the 2P-population always forms centrally-concentrated.      
More recently, \citet{lei23} have shown that the full range of behaviors coexist among dynamically young globular clusters (i.e. clusters whose relaxation time-scale is comparable to their old stellar age), from one population being more centrally-concentrated than the other, to both populations being similarly distributed.  The latter result cannot be due to radial mixing owing to the cluster young dynamical age \citep[see also][]{vbek15,hoo21}.

The diversity of behaviors observed among dynamically young globular clusters justifies the assumption made in this model that 1P- and 2P-stars can be similarly distributed in newly formed clusters, leading to the subsequent constancy over time of their 1P-star fraction.  This is a first step, however, and a generalization with population-dependent mass losses will be presented in a forthcoming paper.  

To assume the constancy over time of $F_{1P}$ implies that aging clusters move solely to the left in (mass, $F_{1P}$) space.  We therefore expect \gcs subjugated to a strong external tidal field to be located to the left of those evolving in gentle tidal fields.  That is, \gcs  orbiting closer to the Galactic center ("inner" clusters) should be located to the left of those with larger Galactocentric distances ("outer" clusters).  On the average, this is indeed observed \citep[compare the grey and red lines in top-left panel of fig.~7 in][]{mil20}.  This shift between inner and outer clusters was previously interpreted as inner clusters standing {\it below} outer clusters in (mass, $F_{1P}$) space \citep{zen19,mil20}.  This was seen as the result of the preferential removal of the more vulnerable 1P-stars from the more tidally-stripped "inner" clusters.  If this is the case, this downward shift is necessarily accompanied by a leftward shift accounting for the corresponding cluster mass loss.  As we shall see in this contribution, however, a pure leftward shift can also explain the respective distributions of inner and outer clusters in (mass,$F_{1P}$) space.              

With dissolving \gcs feeding the stellar field, the attested scarcity of 2P-field stars in the Galactic halo \citep{car10a,mar11} may also have motivated earlier scenarios in which 1P-stars are those preferentially stripped from their natal clusters.  Yet, we shall see that this observational constraint is naturally met by the present model, as a result of the high cluster-mass threshold imposed for 2P-star formation.  \\

The outline of the paper is as follow.  In Sec.~\ref{sec:mod}, we present the model and how it accounts for the successive cluster evolutionary phases, from newly-formed gas-embedded clusters to 12\,Gyr-old ones.  Section~\ref{sec:comp} compares our 12\,Gyr-old model tracks with the observed cluster distribution in (mass,$F_{1P}$) space, while also discussing the impact of cluster primordial mass segregation and of a top-heavy IMF.  In Sect.~\ref{sec:edges}, we show how dynamical friction, the scarcity of clusters at high mass and at large Galactocentric distances curtail the observed distribution.  The respective behaviors of outer and inner \gcs are examined in Sec.~\ref{sec:inout}, while Magellanic Clouds and Fornax clusters are discussed in Sec.~\ref{sec:mag}.  Section~\ref{sec:2PHFS} demonstrates that the model meets the observational constraints of a low 2P-star fraction in the Galactic halo field.  Finally, we present our conclusions, and outline our future work, in Sec.~\ref{sec:conc}.      

\section{Model Philosophy}     
\label{sec:mod}

\noindent Model milestones are as follow.  \\
(1) Forming clusters must reach a threshold mass $m_{th}$ to be polluted in hot hydrogen-burning products,  an hypothesis that builds on the stronger multiple-populations patterns observed in more massive clusters.  In most of this contribution, we adopt $m_{th}=10^6\,\Ms$.  This is of the order of the minimum mass a cluster must achieve to form, via stellar collisions, a Super Massive Star in its dense central regions (\citealt{gie18}; see also \citealt{fre06}).  We shall see, however, that the value of $m_{th}$ is not that rigidly fixed.  \\
(2) Once the pollution of a cluster starts, the entirety of its \sfing gas is polluted.  This may seem an extreme assumption, and we will tone it down in a forthcoming paper.  But for now, it leads to two key model simplifications: \\
\indent (2a) In newly-formed clusters with multiple populations, the mass of the pristine-population is always $m_{th}$, and the pristine mass fraction obeys $F_{1P} = m_{th}/m_{ecl}$, with $m_{ecl}$ the mass of clusters at the end of their formation; \\ 
\indent (2b) As soon as they start forming, 2P-stars form in all cluster regions, including cluster outskirts in case of a centrally-located source of pollution.  As a result, 2P-stars are not more centrally located than 1P-stars, and both groups of stars - pristine- and polluted-ones - are lost with the same probability as clusters age.  In other words, clusters evolve at constant $F_{1P}$.  All we therefore need to model is the cluster mass decrease due to  dynamical evolution, with the same fractional mass loss applied to the whole cluster, its 1P and 2P populations.  \\ 
(3) The formation of clusters (Sec.~\ref{ssec:ecl}) is terminated once their massive stars expel the residual \sfing gas.  Two successive evolutionary phases follow: \\ 
\indent (3a) dynamical response to residual \sfing gas expulsion (i.e.~violent relaxation; Sec.~\ref{ssec:vr}); \\
\indent (3b) long-term secular evolution during which clusters experience stellar evolutionary mass losses, 2-body-relaxation and tidal stripping due to the host-galaxy tidal field (Sec.~\ref{ssec:sec}).   \\

We shall show that cluster models starting with the scaling $F_{1P} = m_{th}/m_{ecl}$ and evolving at constant $F_{1P}$ reproduce remarkably well, at an age of $\simeq 12$\,Gyr, the relation between observed 1P fraction $F_{1P}^{obs}$ and present-day mass $m_{prst}$ of globular clusters. 

\subsection{The Gas-Embedded-Cluster Stage ('Ecl')}
\label{ssec:ecl}
     
Let us consider a star cluster forming out of a clump of gas, with $t=0$ defining star-formation onset.  At $t=t_{th}$, the stellar mass of the forming cluster has reached the threshold $m_{th}$ at which the intra-cluster gas starts being polluted with \hhb products: the formation of 2P-stars starts. 
Let us also assume the pollution of the {\it entire} intra-cluster gas at $t=t_{th}$.  That is, 1P/pristine stars form only when $t < t_{th}$, while 2P/polluted stars are the only ones to form once $t \geq t_{th}$.  As a result, the stellar mass threshold $m_{th}$ equates with the mass of the pristine population, while the mass of the polluted population is the cluster mass formed at $t \geq t_{th}$.  

\begin{figure}
\begin{center}
\includegraphics[width=0.49\textwidth]{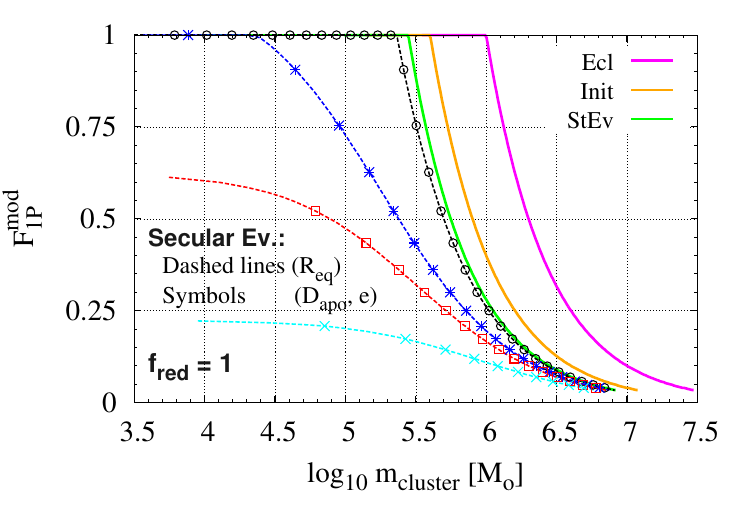}
\caption{Model relations between the mass fraction of pristine stars in clusters $F_{1P}$ and the logarithmic cluster mass.  The solid magenta (rightmost) line shows the relation for embedded clusters (Eq.~\ref{eq:f1p-ecl} with $m_{th}=10^6\,\Ms$).  The solid orange line to its left depicts the relation at secular evolution onset (Eq.~\ref{eq:f1p-init}), with clusters retaining a fixed fraction $F_{bound}^{VR}=0.40$ of their stellar mass by the end of violent relaxation.  The green line shows the effects of 30\,\% stellar evolutionary mass losses.  Symbols and dashed lines show the impact of a 12\,Gyr-long secular evolution in the tidal field of the host galaxy (circular velocity $V_c=220\,km\cdot s^{-1}$).  Symbols correspond to clusters on elliptical orbits of eccentricity $e=0.6$ and apocentric distances $D_{apo}=$ 20.0 (open circles), 3.75 (asteriks), 2.50 (open squares) and 1.25\,kpc ($\times$-symbols).  Dashed lines correspond to circular orbits of radius $R_{eq}=(1-e)D_{apo}=8.0$, 1.5, 1.0 and 0.5 (black, blue, red and cyan dashed lines, respectively).  By definition of $R_{eq}$, such orbits are characterized by the same cluster dissolution time-scale as their elliptical counterparts (see Eq.~\ref{eq:tdiss} and text for details).  } 
\label{fig:mth}
\end{center} 
\end{figure}

With $m_{ecl}$ the stellar mass of the cluster at the end of its formation (i.e. when the residual \sfing gas is expelled), the mass fraction of the pristine population in the cluster obeys 

\begin{equation}
\begin{split}
F_{1P}(m_{ecl})= & \frac{m_{th}}{m_{ecl}} {\rm ~~if~~} m_{ecl}>m_{th} \\
               = & 1 {\rm ~~otherwise}\;.
\label{eq:f1p-ecl}
\end{split}
\end{equation}   
$F_{1P}=1$ corresponds to clusters that have failed to form a stellar mass larger than the threshold $m_{th}$ and, therefore, that are made of 1P-stars only.    
Equation~\ref{eq:f1p-ecl} is depicted for $m_{th} = 10^6\,\Ms$ as the solid magenta (rightmost) line in Fig.~\ref{fig:mth}.  We shall discuss further the value of $m_{th}$ in Sec.~\ref{ssec:vr} and \ref{sec:comp}.

With the whole intra-cluster medium polluted at $t=t_{th}$ already, 2P-populations are, from the start of their formation, spread all through their host clusters.  Therefore, when clusters expand in response to gas expulsion, they lose their 1P- and 2P-stars equally likely, and they evolve at constant $F_{1P}$.       

\subsection{Violent Relaxation: Towards the Initial Stage ('Init')}
\label{ssec:vr}

As massive stars launch their winds and eventually explode as Type~II supernovae, the intra-cluster gas is progressively swept into an outwardly expanding shell \citep[e.g.][]{bbt95,gey01,rah17,kou23}.  This expulsion of the residual star-forming gas out of the cluster terminates its formation\footnote{Unless stars form in the expanding shell and eventually fall back onto the cluster, thereby augmenting its mass, as envisioned by \citet{bbt95}, \citet{par04}, and \citet{rah17}. }.   

Following gas expulsion, clusters expand and lose stars (aka violent relaxation).  The mass of clusters at the end of violent relaxation being also their mass at the onset of their long-term secular evolution, we coin it the cluster "initial mass" $m_{init}$ .  With $F_{bound}^{VR}$ the stellar mass fraction retained by clusters at the end of violent relaxation, we have  
\begin{equation}
m_{init}=F_{bound}^{VR}\cdot m_{ecl}\,.
\label{eq:minit}
\end{equation}   

The cluster bound fraction $F_{bound}^{VR}$ depends on a large array of cluster properties at gas  expulsion: \stf efficiency \citep{hil80, lad84}, virial state \citep{ver90,goo09,smi13,far18}, the time scale $\tau_{GExp}/\tau_{cross}$ on which the \sfing gas is expelled \citep{mat83,lad84,gey01}, star and gas radial density profiles \citep[][]{ada00,par13,shu17}, amount of substructures inside forming clusters \citep[][]{smi13,far18}, the interplay between cluster compactness and external tidal field \citep{goo97,bk07,shu19}.  For the sake of simplicity, we consider a fixed value of the bound fraction and we adopt $F_{bound}^{VR}=0.40$. We discuss our choice further in Appendix~A. 

Since we assume the pristine star fraction $F_{1P}$ to be conserved through cluster evolution, Eqs~\ref{eq:f1p-ecl} and \ref{eq:minit} yield

\begin{equation}
F_{1P}(m_{init})=F_{1P}(m_{ecl})=\frac{F_{bound}^{VR}m_{th}}{m_{init}}\;.
\label{eq:f1p-init}
\end{equation} 

Equation~\ref{eq:f1p-init} defines the relation between the pristine star fraction $F_{1P}$ and the cluster mass $m_{init}$ at the onset of secular evolution.  It is shown, for $F_{bound}^{VR}=0.40$ and $m_{th}=10^6\,\Ms$, as the solid orange line in Fig.~\ref{fig:mth}.  It is the solid magenta line shifted leftwards by ${\rm log}_{10}(F_{bound}^{VR})=-0.4$.  At this stage of cluster evolution, multiple-populations clusters have a mass of at least $F_{bound}^{VR}m_{th}$.
   
We remind the reader that our objectives are: (i) to model the 1P-star fraction in dependence of the cluster mass at an age of 12\,Gyr, and (ii) to compare model results with the available observational data.  The relation $F_{1P}(m_{init})$ obtained above will thus serve as an "anchor" out of which 12~Gyr-old relations unfold.  Its crucial parameter is neither the bound fraction $F_{bound}^{VR}$ alone, nor the stellar mass threshold $m_{th}$ alone.  Rather, this is the {\it product} $F_{bound}^{VR} m_{th}$ that matters since it determines the locus of clusters in $(m_{init},F_{1P})$ space. We shall explain in due time how to adjust the product $F_{bound}^{VR}m_{th}$, therefore tying our respective choices of $m_{th}$ and $F_{bound}^{VR}$.  

\subsection{Stellar-Evolution ('StEv') and Long-Term Secular Evolution ('SecEv'): Towards an age of 12\,Gyr}
\label{ssec:sec}

To model the mass of \gcs at an age of 12\,Gyr, which we refer to as their present-day mass, $m_{prst}$, we need to model their long-term mass losses due to stellar evolution, 2-body relaxation, and tidal stripping.  To this purpose, we use Eqs~10 and 12 of \citet{bm03}.  Their Eq.~10 quantifies the time-span needed for clusters, born with a \citet{kro01} IMF and without primordial mass segregation, to dissolve in their host galaxy.  We refer to this time-span as $t_{diss}^{BM03}$.  Their Eq.~12 provides us with the cluster mass at a given age, given their initial mass and dissolution time-scale.  The host galaxy is described as a spherically-symmetric logarithmic potential of circular velocity $V_c$, corresponding to an isothermal sphere in terms of volume density profile.  

We define the cluster dissolution time-scale of our model, $t_{diss}$, as a function of the cluster dissolution time-scale $t_{diss}^{BM03}$ of \citet[][their Eq.~10]{bm03}   
\begin{equation}
\begin{split}
t_{diss} &       = f_{red} ~ t_{diss}^{BM03} \\
         & \propto f_{red} ~ f(m_{init}, x, \beta) ~  \frac{(1-e)D_{apo}}{V_c}\;.
\label{eq:tdiss}
\end{split}
\end{equation}
$f_{red}$ is an additional factor that we introduce and that we discuss later on.  For now, we set $f_{red}=1$, that is, we use Eq.~10 of \citet{bm03} as it is. 
$V_c$ is the host-galaxy circular velocity, $D_{apo}$ and $e$ the apocentric distance and eccentricity of the cluster orbit.  $m_{init}$ is the cluster initial mass, while $x$ and $\beta$ set the cluster initial concentration, either $(x,\beta)=(0.75,1.91)$ for a King concentration parameter $W_0=5.0$ or $(x,\beta)=(0.82,1.03)$ for $W_0=7$.  $f(m_{init}, x, \beta)$ is thus a function of the cluster initial mass and concentration.   

Equation~12 of \citet{bm03} describes the evolution with time of the mass of clusters, as they experience stellar-evolution mass losses, 2-body relaxation and tidal stripping.  For a canonical IMF \citep[e.g.][]{kro01}, stellar evolutionary mass-losses amount to $\simeq 30$\,\% of the initial cluster mass.  They are assumed to take place instantaneously at the onset of secular evolution, as justified by their short duration ($\simeq 1$\,Gyr for a canonical IMF) compared to the old age of \gcs \citep[$\simeq 12$\,Gyr,][]{vdb13}.  Thereafter, the cluster mass decreases linearly with time due to combined internal (2-body relaxation) and external (tidal stripping) effects, until the cluster gets completely dissolved at an age given by $t_{diss}$.  
 
With Eqs~10 and 12 of \citet{bm03}, we are equipped to evolve the relation $F_{1P}(m_{init})$ up to an age of 12\,Gyr.  In Fig.~\ref{fig:mth}, the solid green line shows the impact of stellar evolutionary mass losses, that is, it corresponds to $F_{1P}(m_{init})$ -- orange line in Fig.~\ref{fig:mth} -- shifted leftwards by $F_{StEv}=70$\,\%.   

When estimating the cluster dissolution time-scale, we assume a circular velocity $V_c=220\,km\cdot s^{-1}$ for the Milky-Way-like host galaxy and a cluster initial concentration $W_0=5$ (i.e. $x=0.75$ and $\beta=1.91$).  A 12\,Gyr-old track then corresponds to a given set $(D_{apo}, e)$.  The more eccentric the orbit and/or the smaller the apocentric distance $D_{apo}$ (i.e. the smaller the product $(1-e)D_{apo}$ in Eq.~\ref{eq:tdiss}), the stronger the mean external tidal field experienced by clusters, the shorter their dissolution time-scale $t_{diss}$, and the smaller the mass fraction they retain by an age of 12\,Gyr.  Additionally, for a given orbit $(D_{apo}, e)$, the lower the cluster initial mass $m_{init}$ (i.e. the higher $F_{1P}$), the shorter the cluster dissolution time-scale, and the greater the cluster mass fraction lost through secular evolution.  Therefore, the leftward shift separating 12\,Gyr-old tracks $F_{1P}(m_{prst})$ from the initial one $F_{1P}(m_{init})$ is larger for smaller cluster masses, lower apocentric distances, and/or larger orbital eccentricities.   We remind the reader that a constant $F_{1P}$ is assumed all through the cluster evolution.  That is, in Fig.~\ref{fig:mth}, each point of the magenta/orange/green solid lines experiences a horizontal leftward shift to the next evolutionary stage. 

Symbols in Fig.~\ref{fig:mth} depict the 12\,Gyr-old tracks for $e=0.6$ and $D_{apo}=20.0$, 3.75, 2.50 and 1.25\,kpc (open circles, asterisks, open squares and $\times$-symbols, respectively).  $e=0.6$ is the median orbital eccentricity of our \gc sample \citep[i.e. \gcs with $F_{1P}$ estimates; based on the apo- and pericentric distances of the cluster orbits computed by][]{bau19}.  The cluster sample is presented in Sec.~\ref{sec:comp}.

Since the dissolution time-scale of a cluster scales as $t_{diss} \propto (1-e)D_{apo}$ (Eq.~\ref{eq:tdiss}), we can define an {\it equivalent orbital radius $R_{eq}=(1-e)D_{apo}$} as the radius of the circular orbit ($e=0$) that yields the same cluster dissolution time-scale as for an orbit of apocentric distance $D_{apo}$ and eccentricity $e$.  With $e=0.6$ and $D_{apo}=20.0$, 3.75, 2.50 and 1.25\,kpc, the corresponding equivalent radii are $R_{eq}=8.0$, 1.5, 1.0, and 0.5\,kpc.  Their respective 12\,Gyr-old tracks are shown as the dashed lines in Fig.~\ref{fig:mth}.  Per definition of $R_{eq}$, they coincide with the symbols of identical colors.  In other words, the locus of a track depends neither on the apocentric distance alone, nor on the cluster orbital eccentricity alone.  When quantifying the cluster dissolution time-scale, this is the product $(1-e)D_{apo} = (1-0)R_{eq}$ that matters.  Two initially identical clusters with different orbital eccentricities $e$ and apocentric distances $D_{apo}$ evolve to the same locus in $(m_{prst},F_{1P})$ space provided that their product $(1-e)D_{apo}$ is the same (e.g. the larger apocentric distance of a cluster compensates for its higher orbital eccentricity). 
This will constitute a great advantage when comparing models and observations: there is no need to compare model tracks with the locus of \gcs of similar eccentricities.  This is in terms of $R_{eq}=(1-e)D_{apo} $ that model tracks and observational data points are to be compared.

\begin{figure}
\begin{center}
\includegraphics[width=0.49\textwidth]{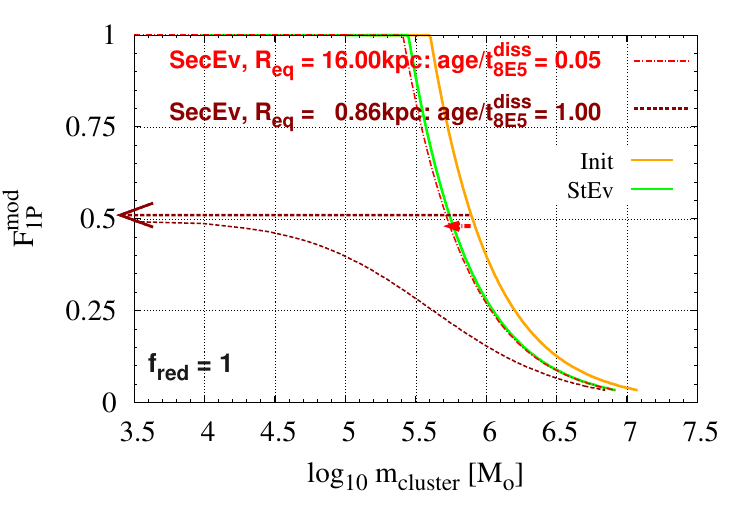}
\caption{Mass loss experienced by a 12\,Gyr-old cluster with $m_{init}=8\cdot 10^5\,\Ms$ ($F_{1P}=0.5$), for two equivalent radii $R_{eq}=(1-e)D_{apo}= 16.00$ and 0.86\,kpc (dash-dotted red and dashed dark-red arrows, respectively), $W_0=5$ and $V_c=220\,km\cdot s^{-1}$.  Clusters with $F_{1P}>0.5$ ($F_{1P}<0.5$) are initially less (more) massive than $m_{init}=8\cdot 10^5\,\Ms$.  They experience therefore greater (smaller) mass losses (see the tracks with the same line- and color-codings as the arrows).   The track for $R_{eq}=0.86$\,kpc stretches horizontally towards low masses, indicating that a cluster with $m_{init}=8\cdot 10^5\,\Ms$ is on the edge of dissolution at an age of 12\,Gyr when $R_{eq}=0.86$\,kpc.   Ratios between the cluster age and dissolution time-scale are quoted at the top of the figure.  Solid orange and green lines as in Fig.~\ref{fig:mth}.   }
\label{fig:tdiss}
\end{center} 
\end{figure}   
 
Figure~\ref{fig:tdiss} visualises with horizontal arrows the leftward shifts that a 12Gyr-old cluster with $F_{1P}=0.5$ has experienced for $R_{eq}=$ 16.00 and 0.86\,kpc.  $F_{1P}=0.5$ implies $m_{ecl}=2\cdot10^6\,\Ms$ (with $m_{th}=10^6\,\Ms$) and $m_{init}=8\cdot10^5\,\Ms$ (with $F_{bound}^{VR}=0.4$).  The ratio between cluster age (12\,Gyr) and dissolution time-scale is quoted at the top of the figure, assuming $W_0=5$ and $V_c=220\,km\cdot s^{-1}$.  When $R_{eq}= 16.0$\,kpc (outer-halo orbit/weak tidal field; dash-dotted red line), the cluster age amounts to a minor fraction of its dissolution time-scale, and the lost mass does not differ much from the stellar evolutionary mass losses.  That is, the 12\,Gyr-old track almost overlaps with the light-green post-stellar-evolutionary mass-loss track, especially at high mass.  When $R_{eq}= 0.86$\,kpc (strong tidal field; dashed dark-red line), the cluster age equates with its dissolution time and, therefore, the cluster mass tends towards 0. 
 
\begin{figure}
\includegraphics[width=0.49\textwidth, trim = 0 55 0 0]{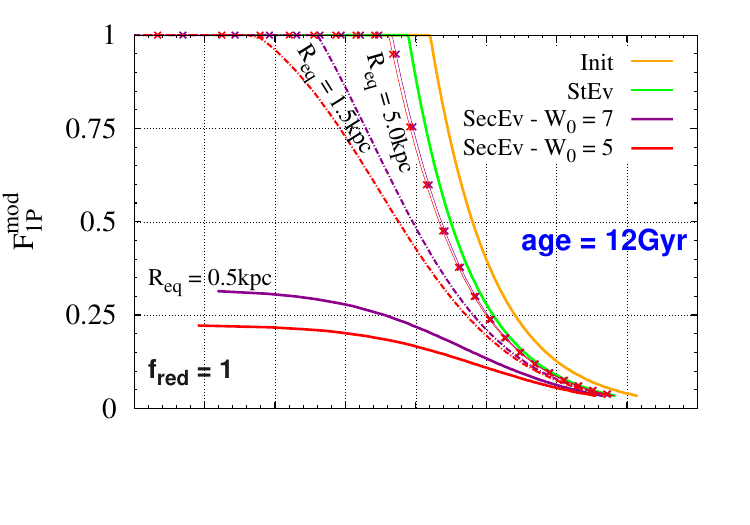}\\
\includegraphics[width=0.49\textwidth, trim = 0  0 0 0]{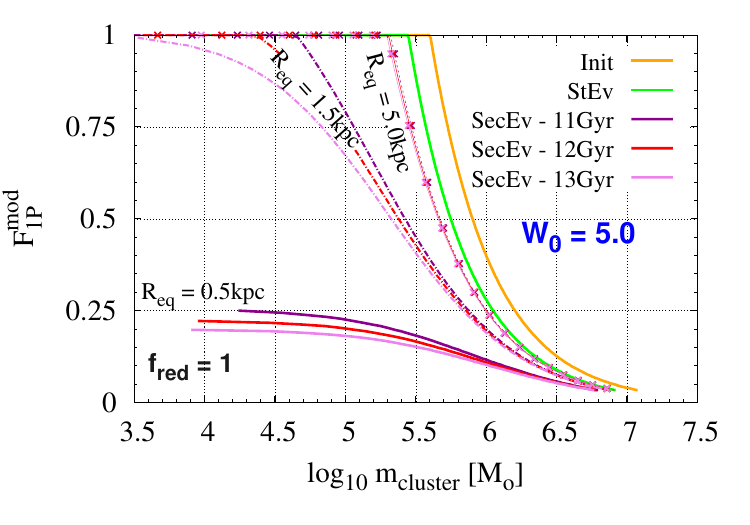}
  \caption{Top panel: Impact of the initial cluster concentration $W_0$ on 12\,Gyr-old model tracks in (mass,$F_{1P}$) space.  Bottom panel: Impact of cluster age on model tracks in (mass,$F_{1P}$) space, for a fixed initial cluster concentration $W_0=5$.  Lower initial cluster concentrations and/or older ages move tracks leftwards owing to the greater cluster mass losses they induce.}
 \label{fig:disc} 
\end{figure} 

Massive clusters at short distances from the Galactic centre are also affected by dynamical friction, the impact of which is considered in Sec.~\ref{ssec:dynfric}.       

Figure~\ref{fig:disc} illustrates how varying the initial concentration $W_0$ (top panel) and the age (bottom panel) of \gcs affect 12\,Gyr-old tracks.  Equivalent radii $R_{eq}=$ 0.5, 1.5 and 5.0\,kpc are considered. For a given age (12\,Gyr, top panel), a lower initial concentration yields a greater amount of mass-loss, and the tracks with $W_0=5$ (in red) are thus located leftward to those with $W_0=7$ (in violet).  \citet{bm03} show that low-concentration globular clusters, namely, clusters starting with $W_0=3$, dissolve early in their evolution.  For a given set of parameters $(V_c, D_{apo}, e)$ and for the $W_0$ values tested by \citet{bm03}, the tracks with $W_0=5$ therefore define a lower limit in $(m_{prst},F_{1P})$ space.  For a fixed initial concentration ($W_0=5$, bottom panel), older cluster ages are conducive to greater fractional mass losses and, thus, larger leftward shifts in $(m_{prst},F_{1P})$ space.  Note that the impact of varying the initial concentration ($W_0=5$ or $W_0=7$) or the age of \gcs (age=11, 12 or 13\,Gyr) is negligible compared to variations of the equivalent orbital radius $R_{eq}=(1-e)D_{apo}=(1+e)D_{peri}$.   

Figures~\ref{fig:mth}-\ref{fig:disc} show that a Hubble-time of dynamical evolution in the Galactic tidal field segregates \gcs in ($m_{prst}$, $F_{1P}$) space according to their Galactocentric distance, an aspect that we shall discuss further in Sec.~\ref{sec:inout}.

\begin{figure}
\begin{center}
\includegraphics[width=0.49\textwidth]{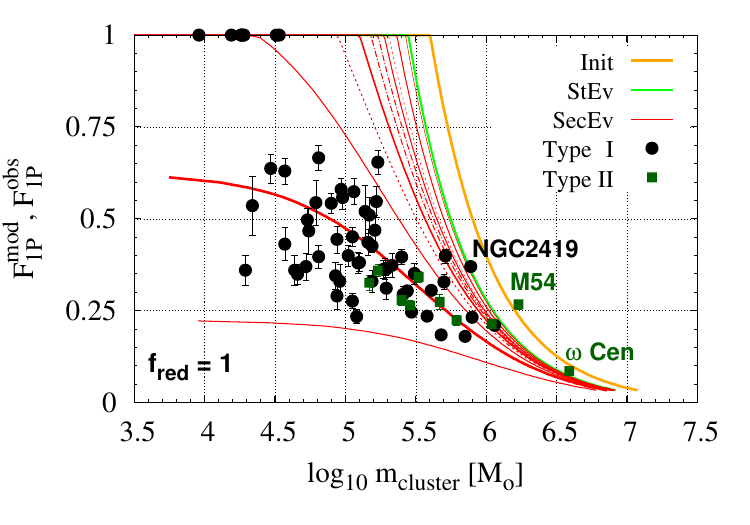}
\caption{Pristine-star fraction $F_{1P}$ in dependence of cluster masses.  Plain symbols depict the observed number fraction $F_{1P}^{obs}$ of Galactic \gcs  (both Type~I and Type~II) in dependence of their present-day mass $m_{prst}$.  The orange line is the model initial sequence $F_{1P}^{mod}(m_{init})$ (Eq~\ref{eq:f1p-init} with $F_{bound}^{VR}m_{th}=4 \cdot 10^5\,\Ms$).  Red lines are 12Gyr-old tracks with $W_0=5$, $V_c=220\,km\cdot s^{-1}$ and, from top to bottom, $R_{eq}=$30.0, 8.0, 5.0, 4.2, 3.5, 3.0, 2.5, 2.0, 1.5, 1.0 and 0.5\,kpc.  Short-dashed lines depict the tracks for $R_{eq}=$2.0 and 5.0\,kpc.  Dash-dotted lines depict the tracks for $R_{eq}=$3.0 and 3.5\,kpc.
}
\label{fig:nofred}
\end{center} 
\end{figure}   

\section{Comparison with Observations: First Step}
\label{sec:comp}

Figure~\ref{fig:nofred} compares in ($m_{prst}$, $F_{1P}$) space the 12Gyr-old model tracks with the observations available for Galactic globular clusters.  We denote observed and modeled pristine-star fractions as $F_{1P}^{obs}$ and $F_{1P}^{mod}$, respectively.  Note that $F_{1P}^{mod}$ are stellar {\it mass} fractions (see Eq.~\ref{eq:f1p-ecl}), while $F_{1P}^{obs}$ are star {\it number} fractions inferred from \gc chromosome maps \citep[see e.g.~Sec.~4.2 in ][for details]{mil17}.  Additionally, model mass fractions $F_{1P}^{mod}$ refer to the cluster {\it total} mass, while  measured number fractions $F_{1P}^{obs}$ stem from a limited sample of observed stars.  When comparing model tracks and observational data points, we therefore implicitly assume that model stellar mass fractions $F_{1P}^{mod}$ and measured number fractions $F_{1P}^{obs}$ are equivalent.  

Measured number fractions $F_{1P}^{obs}$ are taken from Table~3 in \citet{don21}, which summarizes and updates the results obtained by \citet{mil17}, \citet{mil18} and \citet{zen19}.  We supplement this data set with the results of \citet{jan22} for 5  more \gcs (NGC1904, NGC4147, NGC6712, NGC7006, and NGC7492; see their table~4).
  
Globular clusters Ruprecht~106, Terzan~7, AM~1, Eridanus, Palomar~3, Palomar~4, Palomar~14 and Pyxis are considered single-population clusters \citep[i.e. $F_{1P}^{obs}=1$, ][and references therein]{mil20}.  Present-day cluster masses are taken from \citet{bau19}\footnote{\url{https://people.smp.uq.edu.au/HolgerBaumgardt/globular/}}.  
Our sample consists of 72 Galactic globular clusters, 62 Type~I and 10 Type~II clusters, depicted in Fig.~\ref{fig:nofred} as plain black circles and plain green squares, respectively.
Compared to Type~I clusters, Type~II clusters present in their chromosome maps additional red sequences, possibly the consequence of an internal metallicity spread \citep{mil17}.  

As previously, model tracks at an age of 12\,Gyr have been obtained with $m_{th}=10^6\,\Ms$, $F_{bound}^{VR}=0.40$, $F_{StEv}=0.70$, $W_0=5$ and $V_c=220\,km\cdot s^{-1}$.  The initial track $F_{1P}^{mod}(m_{init})$ (orange line) constitutes the anchor out of which 12\,Gyr-old tracks unfold.  To set this anchor, we need to set the product $F_{bound}^{VR} m_{th}$ (Eq.~\ref{eq:f1p-init}).  For this purpose, we use the massive and remote \gc NGC2419.  NGC2419 has a present-day mass $m_{prst} \simeq 8 \cdot 10^5\,\Ms$, a Galactocentric distance of 95.93\,kpc, and its peri- and apocentric distances are $D_{peri}=16.52$\,kpc and $D_{apo}=90.96$\,kpc, 
respectively\footnote{\url{https://people.smp.uq.edu.au/HolgerBaumgardt/globular/combined\_table.txt}}.   Its  equivalent radius is thus $R_{eq}=(1-e)D_{apo}\simeq 28$\,kpc.  Its pristine star fraction is $F_{1P}^{obs}=0.37$ \citep{zen19}.  
Owing to both its large mass and the weak tidal field in which it evolves, NGC2419 must have experienced little tidal stripping.  Its current pristine-star fraction must be similar to the initial one, and its current mass must be close to the initial one minus the stellar-evolutionary mass losses (assuming those have actually been lost from the cluster potential).  We therefore adjust the product $F_{bound}^{VR} m_{th}$ such that NGC2419 settles on the post-stellar-evolution mass-losses (green) track (see Fig.~\ref{fig:nofred}).  That is, $F_{bound}^{VR} \times m_{th} = F_{1P} \times m_{init} \simeq F_{1P} \times m_{prst}/F_{StEv} = 0.37 \times 8\cdot10^5 / 0.70 = 4\cdot10^5\,\Ms$.  The threshold stellar mass $m_{th}$ that a forming cluster must reach to form 2P stars and the post-violent-relaxation bound fraction $F_{bound}^{VR}$ that we have assumed above obey this criterion (i.e. $m_{th}=10^6\,\Ms$, $F_{bound}^{VR}=0.40$).   Note that twice as large a bound fraction combined to twice as low a stellar mass threshold would be conducive to the same initial sequence/orange line and thus to the same 12Gyr-old/red tracks.  The key parameter here is the product $F_{bound}^{VR} m_{th}$.  

$F_{bound}^{VR} m_{th}$ fully defines $F_{1P}^{mod}(m_{init})$ (see Eq.~\ref{eq:f1p-init}), which, as the initial track, represents the high-mass limit of a permitted area (orange line in Fig.~\ref{fig:nofred}): no clusters should be located to the right of it.  Two emblematic Type II globular clusters, NGC6715 (M54) and NGC5139 ($\omega$~Cen), sit on its very edge, which may not be surprising since they are the tidally-stripped nuclei of dwarf galaxies accreted by the Milky Way \citep[][respectively]{iba94,hil00}.  That is, their formation as part of larger structures may have moved them to higher masses compared to genuine globular clusters.   This raises questions regarding the other 8 Type~II clusters.  Can we include them in our sample?  In other words, is their formation process similar to that of Type~I clusters?  For instance, could Type~II cluster progenitors simply be more compact than those of Type~I clusters, thereby providing them with a deeper potential well, hence the ability to retain part of their Type~II Supernova ejecta?  This would yield  the metallicity spreads that we observe today in Type~II clusters \citep[e.g.][]{yon14,bai19,mun21}.  Or, is their formation mechanism fundamentally different?  Do Type~II clusters form as part of dwarf galaxy cores, or are they the merger products of Type~I clusters with different metallicities \citep{bek16}? In the latter cases,  their location in ($m_{prst}$, $F_{1P}$) space cannot be directly compared to our model tracks (our model does not account for cluster mergers, or for the presence around clusters of stars formed in their natal dwarf galaxy).  With the exception of $\omega$ Cen and M54, no firm conclusion regarding the origin of Type~II clusters can be drawn.  All our $F_{1P}$~vs.~cluster-mass diagrams therefore show both Type~I and Type~II clusters, albeit depicted differently, and we note that to include the ten Type II clusters of \citet{mil17} in our sample, or not, will not affect our conclusions.

Having defined the initial relation $F_{1P}^{mod}(m_{init})$, 12~Gyr-old tracks can now be calculated and compared with the locus of globular clusters in $(m_{prst}, F_{1P})$ space.  Figure~\ref{fig:nofred} shows that the regions occupied by model tracks and data points overlap with each other, at least for $R_{eq} \leq 2.0$\,kpc.  In particular, our model explains why the data covers a wider range of $F_{1P}^{obs}$ values at low cluster masses than at high cluster masses.  This is because the fractional mass loss of clusters is a much more sensitive function of the equivalent radius $R_{eq}$ at low mass than at high mass. 

There is one aspect, however, for which the model performs poorly.  The model track going through the cloud of data points has $R_{eq}=1$\,kpc.  In contrast, the median value of $R_{eq}$ for all \gcs included in Fig.~\ref{fig:nofred} is 3.3\,kpc (should we exclude the ten Type~II \gcs from the sample, the median equivalent radius changes only slightly, $R_{eq}=3.4$\,kpc, and the median eccentricity remains $e=0.6$).  Here, we have used the apo- and pericentric distances of the cluster orbits computed by \citet{bau19} to infer their eccentricity $e=(D_{apo}-D_{peri})/(D_{apo}+D_{peri})$ and equivalent radius $R_{eq}=(1-e)D_{apo}$.  
To bring the equivalent radii of model tracks and observed clusters in better agreement, one needs to give an additional leftward push to the tracks.  That is, for a given $R_{eq}$, one needs to hasten the dissolution of clusters, while keeping $F_{1P}$ and the cluster age constant.  We therefore multiply the cluster dissolution time-scale of \citet[][their Eq.~10]{bm03} by a factor $f_{red}$ smaller than unity, which we coin the reduction factor (see our Eq.~\ref{eq:tdiss}).  Physically, it allows us to account for effects shortening the cluster life-expectancy not taken into account by \citet{bm03}.  Such effects could be a top-heavy IMF, or cluster primordial mass segregation. 

\begin{figure}
\begin{center}
\includegraphics[width=0.49\textwidth]{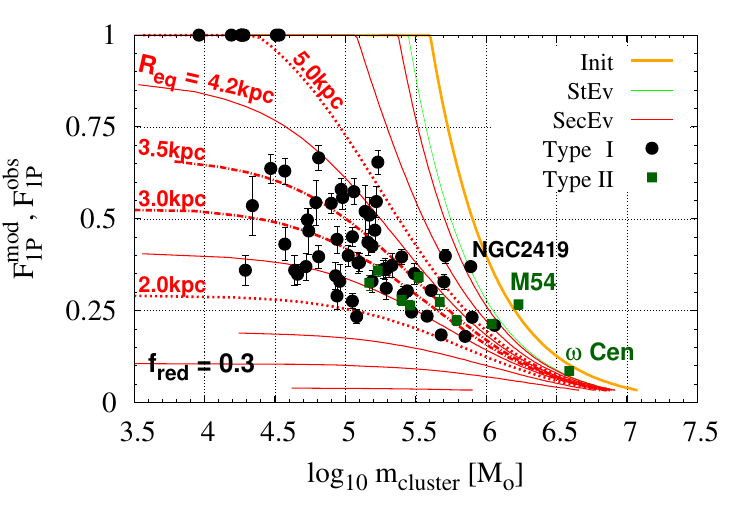}
\caption{Same as Fig.~\ref{fig:nofred}, but with the cluster dissolution time-scale of \citet[][their Eq.~10]{bm03} shortened by a factor 0.3, i.e $f_{red}=0.3$ in Eq.~\ref{eq:tdiss}.  
}
\label{fig:fred}
\end{center} 
\end{figure}   

In Fig.~\ref{fig:nofred}, the track $R_{eq}=1$\,kpc divides the cluster sample into two sub-samples of about the same size (34 clusters out of 72 are located above it).  We thus adopt it as our median track and we use it to calibrate $f_{red}$.  This track must now correspond to $R_{eq}=3.3$\,kpc, the median value of $R_{eq}$ for all data points in Fig.~\ref{fig:nofred}.  Given that $t_{diss} \propto f_{red} (1-e) D_{apo} \propto f_{red} R_{eq}$ (Eq.~\ref{eq:tdiss}), we need $f_{red}=1/3.3=0.30$.  
   
Figure~\ref{fig:fred} shows how the 12\,Gyr-old tracks of Fig.~\ref{fig:nofred} get shifted as a result of the reduced cluster dissolution time-scale.  With $f_{red}=0.30$, the tracks with $R_{eq}=$3.0 and 3.5\,kpc (dash-dotted red lines) now go through the cloud of points, in agreement with the median value of $R_{eq}=3.3$\,kpc for the data.          

Does it make sense to multiply the cluster dissolution time-scale of \citet{bm03} by a factor $f_{red}=0.30$?  We now explore the respective impacts of a top-heavy IMF (Sec.~\ref{ssec:thIMF}), of primordial mass segregation (Sec.~\ref{ssec:mseg}), and, briefly, of a different initial relation $F_{1P}(m_{init})$ (Sec.~\ref{ssec:dif-init}).

\subsection{Top-heavy IMF}
\label{ssec:thIMF}

\citet{bm03} assume for their $N$-body simulations a \citet{kro01} IMF \citep[i.e., for stellar masses higher than $0.5\,\Ms$, the IMF slope is Salpeter-like;][]{slp55}.  Yet, dense stellar systems may possess top-heavy IMFs, e.g. the Arches cluster in the vicinity of the Galactic center \citep{cln12,hos19}, the young nuclear cluster at the center of our Galaxy \citep{lu13}, and 30~Dor in the LMC \citep{sch18}.  A possible reason for top-heavy IMFs emerging from high-density environments is that pre-stellar cores there are so closely packed that they coalesce with each other, thereby increasing their mass and the mass of the stars they give birth to \citep[][]{dib07}.  Such conditions may thus apply to forming globular clusters, although the topic remains debated \citep{che23}.
   
\subsubsection{Cluster present-day mass as a function of IMF top-heaviness}
\label{sssec:mhs20}

To map how the IMF top-heaviness impacts the $F_{1P}(m_{prst})$ relations, we use Eq.~7 in \citet{hag20}.  It  quantifies the cluster dissolution time-scale as a function of the embedded-cluster mass $m_{ecl}$ and of $-\alpha_3$, the logarithmic slope of the IMF $dN_* \propto m_*^{-\alpha_3} dm_*$ for stellar masses $m_*>1\,\Ms$.  In what follows, we denote the cluster dissolution time-scale of \citet{hag20} as $t_{diss}^{HS20}(m_{ecl})$, to make it distinct from the dissolution time-scale $t_{diss}^{BM03}(m_{init})$ of \citet{bm03}.  Figure~4 and Eq.~7 of \citet{hag20} show that $\log (t_{diss}^{HS20}) \propto \log m_{ecl}$, the intercept heavily depending on $\alpha_3$, with cluster lifetimes plummeting for more top-heavy IMFs (see also their Fig.~5 and Table~1).

The cluster dissolution time-scales defined by \citet{hag20} and \citet{bm03} differ on two points: \\  
(1)~\citet{hag20} define their dissolution time-scale as a function of $m_{ecl}$ (the cluster mass at the onset of violent relaxation), while \citet{bm03} define it in dependence of $m_{init}$ (the cluster mass at the end of violent relaxation); \\ 
(2)~$t_{diss}^{HS20}(m_{ecl})$ includes the violent relaxation phase, while $t_{diss}^{BM03}(m_{init})$ does not (i.e. their respective starting points are the embedded stage 'ecl' and the initial stage 'init').  

Yet, Fig.~4 of \citet{hag20} shows that, when $\alpha_3=2.3$, $t_{diss}^{HS20}(m_{ecl})$ and $t_{diss}^{BM03}(m_{init})$ coincide with each other for $m_{ecl} > 2 \cdot 10^5\,\Ms$ (see their red line with plain squares and dash-dotted symbol-free line).  This suggests that to implement $t_{diss}^{HS20}(m_{ecl}, \alpha_3)$ in our model, we could simply swap {\it (i)} the embedded-cluster mass $m_{ecl}$ for the initial cluster mass $m_{init}$, and {\it (ii)} the dissolution time-scale $t_{diss}^{BM03}(m_{init})$ of \citet{bm03} for that of \citet{hag20}, $t_{diss}^{HS20}(m_{ecl})$.  This is indeed the case, for two reasons. 
Firstly, the violent relaxation of massive clusters represents a small fraction of their total lifetime.  As a result, whether the life-expectancy of massive clusters is defined from the onset or from the end of violent relaxation matters little: both dissolution time-scales are indeed comparable, i.e. $t_{diss}^{HS20} \simeq t_{diss}^{BM03}$.  Secondly, with \citet{hag20}'s assumptions, massive clusters (say, $m_{ecl}>3\cdot10^5\,\Ms$) experience adiabatic gas expulsion and remain well-shielded against the Galactic tidal field during their violent relaxation.  They thus lose only a minor fraction of their stars following gas expulsion, actually yielding $m_{init}\simeq m_{ecl}$ (equivalently $F_{bound}^{VR}\simeq 1$).  We demonstrate this in greater detail in Appendix~\ref{sec:apB}.

Having shown that the masses of a high-mass cluster at the onset and at the end of violent relaxation are comparable, the next step is to apply stellar evolutionary mass losses to $m_{init}$.  The cluster bound fractions after stellar evolutionary mass losses, $F_{bound}^{StEv}$, depend on $\alpha_3$.  We take them from the bottom-right panel of Fig.~2 in \citet{hag20} ($\alpha_3 = 2.3$: $F_{bound}^{StEv}=0.6$; $\alpha_3 = 1.9$: $F_{bound}^{StEv}=0.35$; $\alpha_3 = 1.7$: $F_{bound}^{StEv}=0.25$; $\alpha_3 = 1.5$: $F_{bound}^{StEv}=0.20$).  Note that for $\alpha_3 = 2.3$, $F_{bound}^{StEv}=0.6$ is slightly lower than the value of \citet{bm03}.  This is due to different upper limits of the stellar mass spectrum: $m_{up}=15\,\Ms$ in \citet{bm03} and $m_{up}=100\,\Ms$ in \citet{hag20}.  

Following stellar evolutionary mass losses, the cluster mass decreases linearly with time due to the combined effects of 2-body relaxation and tidal stripping \citep[see the bottom-right panel of Fig.~2 in ][]{hag20}.  We can therefore write the present-day mass of a cluster as a function of its initial mass $m_{init}\simeq m_{ecl}$ and of the steepness $\alpha_3$ of its IMF: 
\begin{equation}
m_{prst} =  F_{bound}^{StEv}(\alpha_3) \cdot m_{init} \left(1 - \frac{12\,Gyr}{t_{diss}^{HS20}(m_{init},\alpha_3)}\right)\;.
\end{equation}
This equation is akin to Eq.~12 of \citet{bm03}, with the advantage of covering stellar IMFs more top-heavy than the \citet{kro01} IMF.  

\subsubsection{HS20-tracks: comparison with data set and BM03-tracks}
\label{sssec:ths20}

\begin{figure*}
\begin{center}
\includegraphics[width=0.88\textwidth]{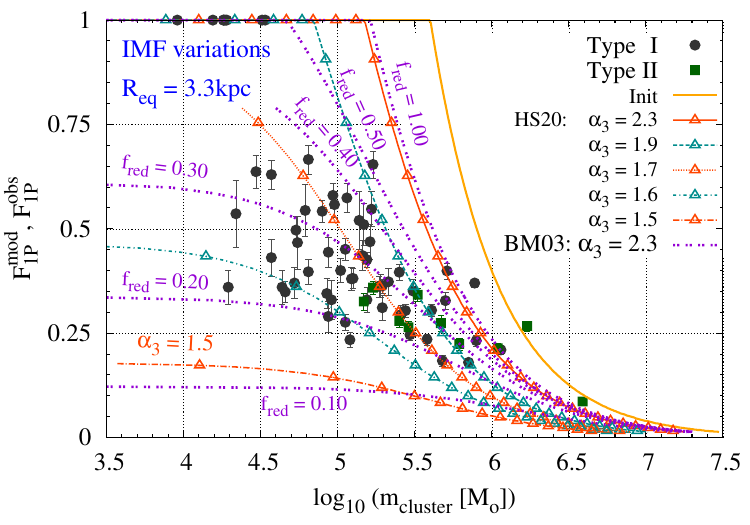}
\caption{How the relation $F_{1P}(m_{prst})$ responds to variations of the cluster IMF steepness $\alpha_3$ (tracks with open triangles, 'HS20', $f_{red}=1$), and to variations of the reduction factor $f_{red}$ (violet tracks, 'BM03', $\alpha_3=2.3$).   To vary the IMF, we use Eq.~7 in \citet{hag20} (see text for details).  The equivalent orbital radius for all models is the median equivalent radius of the data set, i.e. $R_{eq}=3.3$\,kpc.  The initial track $F_{1P}(m_{init})$, Type ~I and Type~II clusters are color/symbol-coded as previously.  The cloud of data points is best bracketted either by BM03-tracks with $f_{red}=0.20{\rm -}0.40$, or by HS20-tracks with $\alpha_3=1.6{\rm -}1.9$.  We nevertheless urge the reader not to immediately conclude, based on this figure, that the IMF of Galactic \gcs is a top-heavy one with $1.6 \lesssim \alpha_3 \lesssim 1.9$ (see Secs~\ref{ssec:mseg} and \ref{ssec:dif-init}).  
}
\label{fig:imf} 
\end{center} 
\end{figure*}

We recall the reader that our objective is to infer the IMF logarithmic slope $-\alpha_3$ best matching the reduction factor $f_{red}=0.3$ applied in Fig.~\ref{fig:fred} to the \citet{bm03} cluster dissolution time-scale $t_{diss}^{BM03}$.  To do so, we will compare tracks obtained with $t_{diss}^{BM03}$ (hereafter BM03-tracks, for which $\alpha_3 = 2.3$, and the varying parameter is $f_{red}$) and with $t_{diss}^{HS20}$ (hereafter HS20-tracks, for which $f_{red}=1$, and the varying parameter is $\alpha_3$).  For both sets of tracks to be fully comparable, they must \\
\indent (a)~be obtained with the same equivalent radius $R_{eq}$.  We take the median orbital radius of our data set $R_{eq}=3.3$\,kpc, such that tracks and data set in $(m_{prst},F_{1P})$ space are also comparable. \\
\indent (b)~unfold from the same initial relation $F_{1P}(m_{init})$ (Eq.~\ref{eq:f1p-init} depicted as the orange line in Figs~\ref{fig:mth}-\ref{fig:imf}), implying that the product $F_{bound}^{VR} \cdot m_{th}$ must be the same. \\
We now make the last adjustments needed to implement (a) and (b). \\

(a)~The median equivalent radius of the data set is $R_{eq}=3.3$\,kpc.  In contrast, \citet{hag20} 's model clusters move on circular orbits at a Galactocentric distance $D_{circ}=8.5$\,kpc.  This implies $R_{eq}=8.5$\,kpc, equivalently a weaker tidal field than what globular clusters experience on the average.  We therefore reduce the cluster dissolution time-scale $t_{diss}^{HS20}$ by a factor $(1-e)=3.3/8.5=0.39$ so as to mimic an elliptical orbit with apocentric distance $D_{apo}=D_{circ}=8.5$\,kpc and eccentricity $e=0.61$ \citep[Eq.~8 in][]{bm03}.  The equivalent radius of the HS20-tracks is now $R_{eq}=(1-e)D_{apo}=(1-e)D_{circ}=3.3$\,kpc, which we will also use for the BM03-tracks.   \\

(b)~Given that $F_{bound}^{VR}=1$ for the HS20-tracks (Sec.~\ref{sssec:mhs20}), we need to lower the corresponding mass threshold down to $m_{th}=4\cdot 10^5\,\Ms$, such that the product $F_{bound}^{VR} \cdot m_{th} = 4\cdot 10^5\,\Ms$ is the same for both the BM03- and HS20-tracks\footnote{
A lower mass threshold $m_{th}$ in case of a top-heavy IMF may actually make sense.  For instance, if hot hydrogen-burning products are given off by an SMS \citep{den14,gie18}, its formation via stellar collisions in its host-cluster central regions -- where stellar collision probability is the highest -- may be accelerated by an excess of massive stars.  This is because stellar collisions involve preferentially massive stars as a result of their large cross-section \citep[see Fig.~22 in][for the build-up of a $\simeq 10^3\,\Ms$-mass star]{fre06} and of their preferential formation in the cluster inner regions where the star-forming gas density is the highest \citep{poly13}.  A top-heavy IMF may thus accelerate SMS formation, thereby lowering the cluster mass threshold $m_{th}$ at which cluster pollution starts.  For now, however, we are unaware of such simulations.  
}.

Figure~\ref{fig:imf} presents the relations $F_{1P}(m_{prst})$ for the cluster dissolution time-scales of  \citet{bm03} and \citet{hag20}.  
BM03-tracks ($\alpha_3=2.3$) are obtained for: $f_{red}=1.00, 0.50, 0.40, 0.30, 0.20, 0.10$.  They are depicted as violet dotted lines along with their respective $f_{red}$ value.  
HS20-tracks ($f_{red}=1$) are obtained for: $\alpha_3 =$ 2.3, 1.9, 1.7, 1.6 and 1.5. They are depicted as lines marked with open triangles (see figure key for their respective $\alpha_3$ value).  

We stress that model tracks in Fig.~\ref{fig:imf} correspond to one given equivalent orbital radius $R_{eq}=3.3$\,kpc combined to either a range of IMF steepnesses $\alpha_3$ along with $f_{red}=1$ (HS20-tracks),  or a range of reduction factors $f_{red}$ along with $\alpha_3=2.3$ (BM03-tracks).  This contrasts with Figs~\ref{fig:nofred}-\ref{fig:fred} in which model tracks correspond to a given IMF ($\alpha_3=2.3$) and a given reduction factor ($f_{red}=1.0$ or 0.3), combined to different equivalent radii $R_{eq}$.  As a sanity check, we check that the BM03-track ($\alpha_3=2.3$) with $f_{red}=1.0$ almost overlaps with the HS20-track ($f_{red}=1.0$) with $\alpha _3 =2.3$ (compare the solid red track with the right-most violet one).

The more top-heavy the IMF, the larger the cluster stellar-evolutionary mass losses.  Clusters with top-heavy IMFs expand therefore more and lose more stars across their tidal boundary than canonical-IMF clusters \citep{hag20}.  Model tracks with $\alpha _3 \leq 2$ are therefore pushed to the left compared to those with $\alpha_3=2.3$.  As an example, clusters with an initial mass $m_{init} \simeq 2.2\cdot 10^6\,\Ms$ (equivalently $F_{1P}\simeq 0.18$) dissolve at an age of 12\,Gyr when $\alpha_3 = 1.5$ (see the dash-dotted red track stretching horizontally at $F_{1P} \simeq 0.18$, thereby indicating cluster dissolution), while they have barely lost half of their mass when $\alpha_3=2.3$ \footnote{Our dissolution time-scale of 12\,Gyr for a cluster with $\alpha_3 = 1.5$ and $m_{ecl}=m_{init} \simeq 2.2\cdot 10^6\,\Ms$ may seem at odds with Fig.~4 of \citet{hag20} where it is $\simeq 30$\,Gyr.  We recall, however, that Fig.~4 of \citet{hag20} assumes cluster circular orbits of radius $D_{circ}=8.5$\,kpc, while the HS20-tracks in Fig.~\ref{fig:imf} assume elliptical orbits of apocentric distance $D_{apo}=8.5$\,kpc and eccentricity $e=0.6$.  The difference between both dissolution time-scales stems from the different orbit eccentricities, i.e. $(1-e)=0.39 \simeq 12\,{\rm Gyr}/30\,{\rm Gyr}$.}. \\

Which $\alpha_3$ values best characterize the cloud of data points, and what are the corresponding reduction factors $f_{red}$ of the \citet{bm03} dissolution time-scale?  
Figure~\ref{fig:imf} shows that the BM03-tracks best bracketting the cloud of data points are those with $f_{red}=0.20 {\rm - } 0.40$ (we thus recover the result of Fig.~\ref{fig:fred}, namely, $f_{red}=0.3$).  In contrast, $f_{red}=0.50$ and $f_{red}=0.10$ have, respectively, too weak and too strong of an impact.  When varying the IMF steepness $\alpha_3$, the cloud of data points is best bracketted by the HS20-tracks with $\alpha_3=1.6$ and $\alpha_3=1.9$.  $\alpha_3 = 1.5$ yields too large cluster mass losses. Figure~\ref{fig:imf} thus suggests that some \gcs form with a top-heavy IMF $1.6 \lesssim \alpha _3 \lesssim 1.9$, although this is not the only possibility to reduce the cluster dissolution time-scale.

\subsection{Primordial mass segregation}
\label{ssec:mseg}
When star clusters are primordially mass segregated \citep[i.e.~massive stars form preferentially in cluster central regions; see e.g.][]{bb06,mye14,xu23}, their inner regions undergo an early excess of stellar evolutionary mass losses (although, contrary to the case of a top-heavy IMF, the cluster as a whole does not).  Because of their central location, such   stellar-evolutionary mass losses remove a greater fraction of the cluster binding energy than for non-primordially segregated clusters.  In turn, this is conducive to greater cluster expansion, the loss of more stars across the tidal boundary and a shorter dissolution time-scale \citep{bau08,ves09,hag14}. 

Based on $N$-body simulations, \citet{hag14} estimate that the dissolution time-scale of strongly primordially-segregated clusters is 4 times shorter than that of non-primordially segregated ones.  This reduction factor is slightly stronger than what we have inferred from Fig.~\ref{fig:fred} ($f_{red}=0.3$).  
Cluster primordial mass segregation is therefore a solution as interesting as a top-heavy IMF for our 12\,Gyr-old model tracks to cover the \gc data points in $(m_{prst},F_{1P})$ space.

\subsection{A different initial relation $F_{1P}(m_{init})$}
\label{ssec:dif-init}

IMF top-heaviness and primordial mass segregation may thus conspire together to reduce the dissolution time-scale established by \citet{bm03} for clusters born without mass segregation and with a Kroupa IMF.  Sections \ref{ssec:thIMF} and \ref{ssec:mseg} show the value $f_{red} \simeq 0.3$ inferred from Fig.~\ref{fig:fred} to be sensible.  Yet, at this stage, we urge the reader not to conclude that \gcs necessarily form with a top-heavy IMF, or mass-segregated.  The reduction factor $f_{red} = 0.3$ also follows from our adopted initial relation $F_{1P} \propto m_{init}^{-1}$ (Eq.~\ref{eq:f1p-init}).  Had we opted for a shallower relation, hence a lower mass threshold $m_{th}$, the ensuing lower initial cluster masses would have called for a less extreme reduction of $t_{diss}^{BM03}$ (i.e.~a higher $f_{red}$) and, therefore, for less top-heavy IMFs and/or less primordial mass segregation.  We shall consider $F_{1P}(m_{init})$ relations different from Eq.~\ref{eq:f1p-init} in a forthcoming paper.

\section{At the inner and outer edges of the Galactic Globular Cluster System}
\label{sec:edges}

In Fig.~\ref{fig:fred}, the density of data points is vanishing towards large Galactocentric distances (i.e. towards the green line), and towards short Galactocentric distances (i.e. towards the bottom-left part of the diagram).  These two effects are expected.
 
Towards large Galactocentric distances, the \gc number density declines steeply \citep[power law of slope $-3.5$; see e.g.][]{djm94,mcl99}.
At short Galactocentric distances, dynamical friction destroys the massive clusters spared by the "classical" tidally-driven dissolution.  For instance, clusters on a stable orbit with $R_{eq}=1.0$\,kpc (equivalent to $D_{apo}=2.50$\,kpc and $e=0.6$) must be more massive than $m_{init}= 4\cdot 10^6\,\Ms$ to survive the Galactic tidal field up to an age of 12\,Gyr (see second lowest red track in Fig.~\ref{fig:fred}).  However, at short Galactocentric distances, such massive clusters are subject to dynamical friction, which causes them to leave their initial orbit and to spiral in towards the Galactic center.  The ever-increasing ambient stellar densities and tidal field they experience eventually tear them apart. 

\subsection{Disruption by Dynamical Friction}  
\label{ssec:dynfric}
\begin{figure}
\begin{center}
\includegraphics[width=0.49\textwidth]{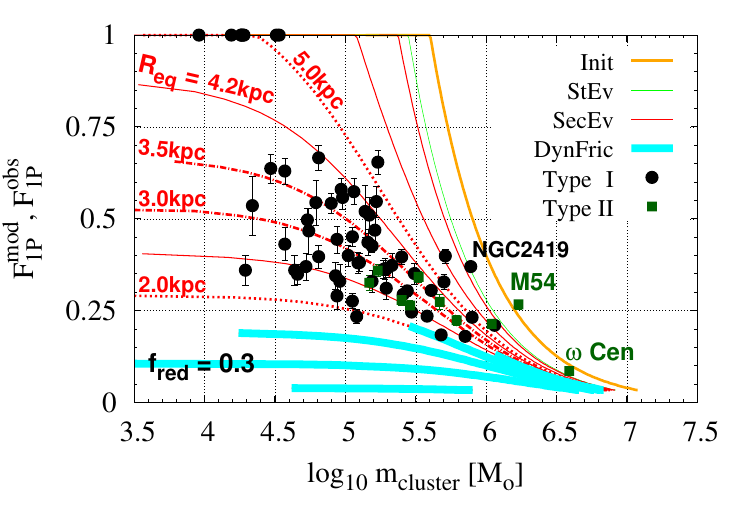}
\caption{Same as Fig.~\ref{fig:fred}, but with thick cyan lines highlighting track portions where \gcs are disrupted within 12\,Gyr by dynamical friction.  Such clusters are initially massive and located at short Galactocentric distances.  Cyan tracks therefore depict regions of the diagram where clusters would survive tidal dissolution for at least 12\,Gyr if tied to a stable orbit ($R_{eq}$ is constant), but that are nevertheless destroyed in 12\,Gyr by dynamical friction.}
\label{fig:fred_fric}
\end{center} 
\end{figure}

The disruption via dynamical friction of a globular cluster initially on a circular orbit of radius $R_{init}$ occurs on a time-scale $t_{fric} \propto R_{init}^2 m_{init}^{-1}$ \citep{bt87,az98}.  That is, the closer to the Galactic center the cluster and/or the higher its initial mass, the faster its disruption via dynamical friction.  
Applying Eq.~7.26 in \citet{bt87}, in which we substitute $R_{init}$ with $R_{eq}$, we find $t_{fric} < 12$\,Gyr if $0.48R_{eq}^2 < m_{init}/10^6\,\Ms$.  In Fig.~\ref{fig:fred_fric}, the segments of the red tracks from Fig.~\ref{fig:fred} satisfying this criterion are highlighted in cyan.  For instance, $R_{eq}=2.0$\,kpc implies that clusters initially more massive than $m_{init} \simeq 2\cdot 10^6\,\Ms$ are destroyed by dynamical friction within 12\,Gyr.  The corresponding red track segment ($m_{prst} > 3 \cdot 10^5\,\Ms$) is thus highlighted in cyan, while the rest of the track, which corresponds to clusters evading dynamical friction owing to their lower initial mass $m_{init}<2\cdot 10^6\,\Ms$, remains untouched. 
Therefore, although clusters with $m_{init}>2\cdot 10^6\,\Ms$ and $R_{eq}=2.0$\,kpc could survive the Galactic tidal field for at least 12\,Gyr, their high initial mass forces them to venture ever closer to the Galactic center (i.e.~$R_{eq}$ decreases), where the stronger tidal field eventually disrupts them.   

Dynamical friction offers therefore an enticing explanation as to why the very bottom-left part of the $(m_{prst},F_{1P})$ space is void of data points (see also Sec.~\ref{ssec:2PHFS_mup}).        

\section{Inner vs. Outer Globular Clusters}
\label{sec:inout}

According to our model, \gcs should be segregated in ($m_{prst}$, $F_{1P}$) space according to their equivalent orbital radius $R_{eq}=(1-e)D_{apo}=(1+e)D_{peri}$.  That is, outer-halo clusters should be located close to the (orange) $F_{1P}(m_{init})$ track, while (surviving) inner-halo clusters should be located in the diagram bottom-left part.   
A segregation as a function of cluster pericentric distance $D_{peri}$ was already noticed by \citet{zen19}.  Their fig.~5 shows that, in ($m_{prst}$, $F_{1P}$) space, clusters with large pericentric distances ($D_{peri} > 3.5$\,kpc) tend to be located above those coming closer to the Galactic center.  Here we go one step further and we investigate how clusters behave as a function of $R_{eq}=(1+e)D_{peri}$.  

\begin{figure}
\includegraphics[width=0.49\textwidth, trim = 0 55 0 0]{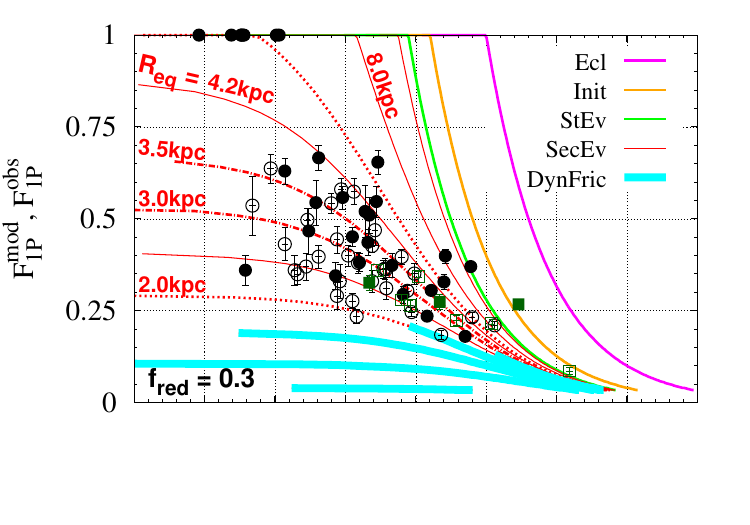}\\
\includegraphics[width=0.49\textwidth, trim = 0 55 0 0]{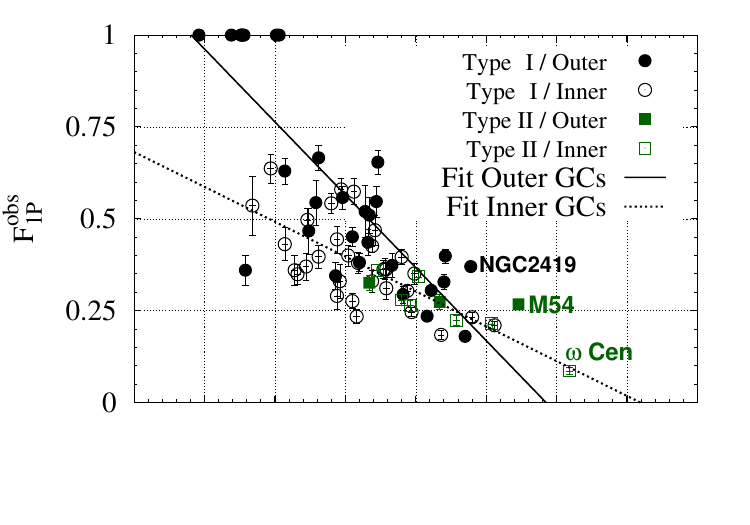}\\
\includegraphics[width=0.49\textwidth, trim = 0  0 0 0]{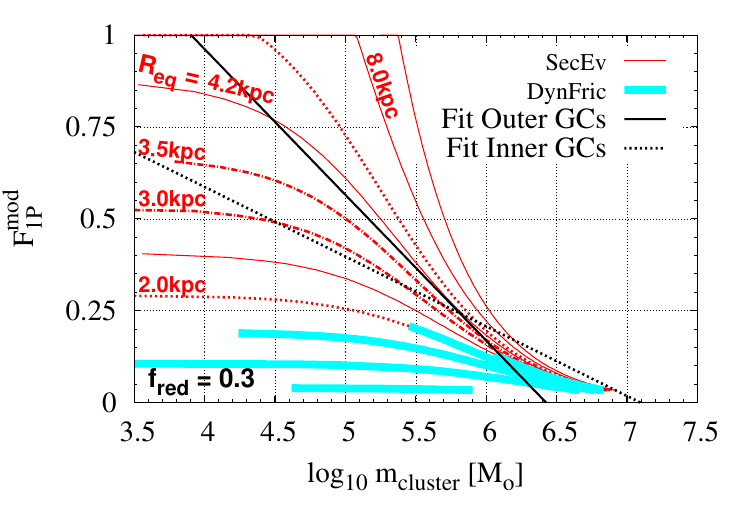}
  \caption{Top panel: Comparison between 12\,Gyr-old model tracks (in red) and cluster data in ($m_{prst}$, $F_{1P}$) space.  Clusters are categorised as either inner clusters ($R_{eq}\leq3.5$\,kpc, open symbols) or outer clusters ($R_{eq}>3.5$\,kpc, plain symbols).  Color- and line-codings are as in Fig.~\ref{fig:fred_fric}.  Middle panel: linear least-squares fits to the inner- and outer-clusters data.  As predicted by the model, the distribution of outer clusters in $(m_{prst},F_{1P})$ space presents a higher intercept and is steeper than that of inner clusters.  Bottom panel: comparison of 12\,Gyr-old model tracks and least-squares fits to the inner- and outer-clusters data. }
 \label{fig:inout} 
\end{figure}

The top panel of Fig.~\ref{fig:inout} depicts clusters with $R_{eq}>3.5$\,kpc as plain symbols ("outer clusters"), and clusters with $R_{eq}<3.5$\,kpc as open symbols ("inner clusters").  The numbers of outer and inner clusters are 33 and 39, respectively (30 and 32 when Type~II clusters are not counted).  With a median orbital eccentricity $e=0.6$, an equivalent orbital radius $R_{eq}=3.5$\,kpc equates with an apocentric distance $D_{apo}=8.75$\,kpc (i.e. about the Solar Circle) and a pericentric distance $D_{peri}=2.2$\,kpc.  The color-coding, for model tracks and observational data, is as in Fig.~\ref{fig:fred_fric}.  On the average, outer clusters tend to be located above inner ones in $(m_{prst},F_{1p})$ space, as already found by \citet{zen19}.  We stress, however, that their interpretation and ours differ.  For \cite{zen19}, the cluster segregation in ($m_{prst}$, $F_{1P}$) space results from the preferential tidal stripping of pristine stars (i.e. downward and leftward shifts combined), as they assume 1P-stars to be the cluster-outskirts dominant population.  In contrast, in our model, inner clusters lose a greater fraction of their initial mass than outer clusters, {\it while not modifying their pristine-star fraction $F_{1P}$} (i.e.~pure leftward shifts).  That is, in the framework of our model, it is more appropriate to state that, in $(m_{prst},F_{1p})$ space, inner clusters (open symbols) tend to be located {\it leftwards} of their outer counterparts (plain symbols) rather than below.  

We note that all clusters with $F_{1P}^{obs}=1$ are outer clusters.  This makes sense since clusters made of 1P-stars only are less massive than those hosting 2P-stars, both in the observed sample and along our initial model track $F_{1P}(m_{init})$.  To survive up to an age of 12\,Gyr, single-population clusters therefore need the weak tidal field of the outer halo.  Our grid of model tracks indeed suggests that such clusters cannot survive unless their equivalent radius $R_{eq}$ is at least $\simeq 4.5$\,kpc.  Among the 8 clusters with $F_{1P}^{obs}=1$, Pal14 has the shortest equivalent radius, $R_{eq}=7.6$\,kpc.  All single-population clusters thus satisfy the model constraint that they survive until today only if their equivalent radius is at least $R_{eq}=4.5$\,kpc. 

\citet{zen19} note that NGC2419 has a fairly large $1P$-fraction compared to other massive clusters (see middle panel of Fig.~\ref{fig:inout}).  Our model allows us to read anew the locus of NGC2419 in ($m_{prst}$, $F_{1P}$) space, which then no longer appears so peculiar.  According to our model, the initial mass of NGC2419 is $m_{init}\simeq 1.1 \cdot 10^6\,\Ms$, corresponding to $F_{1P}=0.37$.  Owing to its remote location in the Galactic halo, it has retained a much greater fraction of its initial mass than other initially more massive clusters located closer to the Galactic center.  As a result, NGC2419 is today one of the most massive globular clusters despite a not-so-low $1P$-fraction.  In contrast, other clusters with similar 1P-fractions ($F_{1P} \simeq 0.37$) have become less massive than NGC2419 owing to the more intense tidal stripping they have endured in more inner regions of the Galactic halo.  

Our grid of model tracks suggests that clusters with large $R_{eq}$ should be distributed inside a region of the diagram that is both steeper and with a higher intercept than clusters with small $R_{eq}$.  Actually, for large $R_{eq}$, the shape of a track mostly mirrors the initial one $F_{1P}(m_{init})$ because of the limited cluster mass losses.  In contrast, at short $R_{eq}$, tracks flatten as a result of cluster dissolution.     
The middle panel of Fig.~\ref{fig:inout} tests whether this effect is indeed present in the observed sample.  The dotted and solid lines represent linear least-squares fits to the inner and outer clusters, respectively, also taking into account outer single-population clusters:  $F_{1P}^{obs,inner}=(-0.19\pm 0.03)\log(m_{prst}^{inner})+(1.35\pm 0.15)$ and $F_{1P}^{obs,outer}=(-0.40\pm 0.05)\log(m_{prst}^{outer})+(2.55\pm 0.24)$. 
The data behave therefore as predicted by the model, outer clusters presenting a steeper relation $F_{1P}(m_{prst})$, and a greater intercept, than inner clusters.  Excluding the ten Type~II clusters from the sample does not modify significantly the fit equations: $F_{1P}^{obs,inner}=(-0.18\pm 0.04)\log(m_{prst}^{inner})+(1.32\pm 0.21)$ and $F_{1P}^{obs,outer}=(-0.42\pm 0.05)\log(m_{prst}^{outer})+(2.65\pm 0.27)$.

Despite inner clusters being located, on the average, leftwards of the outer ones, both groups overlap significantly.  Model tracks are neatly ordered according to $R_{eq}$ because fixed values have been assumed for the reduction factor $f_{red}$ (i.e. $f_{red}=0.3$ in Eq.~\ref{eq:tdiss}), the initial  cluster concentration $W_0$ (i.e. $W_0=5$) and the cluster age (12\,Gyr).  The situation for observed clusters is of course different.  Two clusters with identical and constant $R_{eq}$ are likely to be different initially, e.g.~different initial concentrations $W_0$, different stellar IMFs and/or degrees of primordial mass segregation (Sec.~\ref{ssec:thIMF} - \ref{ssec:mseg}).  They may also experience different environmental conditions as they orbit the Galaxy, due to e.g. different orbit inclinations with respect to the Galactic disk.  

Additionally, accreted \gcs firstly evolve in their natal dwarf galaxies, the tidal field of which is on average weaker than that of the Milky Way.  When accreted late, such clusters should thus present a longer life-expectancy than in-situ clusters.   Finally, some clusters may have retained preferentially $1P$ or $2P$ stars, thereby implying that models do not always evolve at constant $F_{1P}$.  That inner and outer clusters partly overlap in ($m_{prst}$, $F_{1P}$) space -- on top of distinct behaviors on average -- is therefore unsurprising.  \\

The bottom panel of Fig.~\ref{fig:inout} compares the location of model tracks with the least-squares fits to the inner- and outer-clusters data.

\section{Star Clusters of the Magellanic Clouds and Fornax}
\label{sec:mag}

\begin{figure}
\begin{center}
\includegraphics[width=0.49\textwidth]{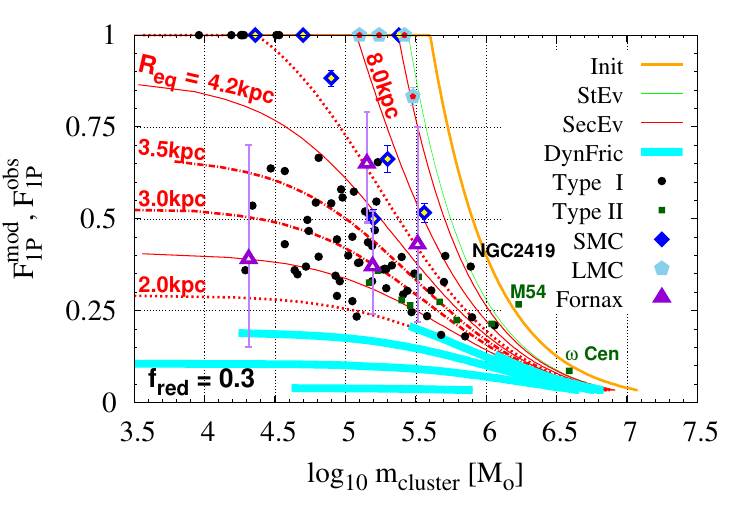}
\caption{Same as Fig.~\ref{fig:fred_fric}, but with the data of the LMC, SMC and Fornax clusters added.  LMC and SMC clusters are located in the outer-halo region of the diagram, as expected for intermediate-age clusters having evolved in the weak tidal field of dwarf galaxies.  The location of the Fornax clusters is somewhat more ambiguous due to the size of the error bars.  For the sake of clarity, the symbols of the SMC/LMC/Fornax clusters are highlighted with central yellow/red/beige symbols, respectively.}
\label{fig:mf1p_mag}
\end{center} 
\end{figure}   

To expand the previous section, we now focus on the star clusters of the Magellanic Clouds.  \citet{mil20}  inferred the fraction of pristine stars in seven clusters of the SMC and in 4 clusters of the LMC.  They are listed in their Table~2, along with the cluster age and present-day mass estimates \citep[for NGC416 and NGC1978, we use the updated pristine-star fractions provided in Table~3 of][]{don21}.  Out of these 11 clusters, 5 have a $1P$-fraction less than unity, that is, they host multiple populations.  They are, in the SMC, Lindsay~1, NGC~121, NGC~339, and NGC~416, and, in the LMC,  NGC~1978.  Owing to their younger age and weaker tidal field they experience, LMC and SMC clusters are less dynamically evolved than their Galactic counterparts.  In $(m_{prst},F_{1P})$ space, we therefore expect to find them fairly close to the orange line depicting the initial track $F_{1P}(m_{init})$.

The locus of LMC and SMC clusters in $(m_{prst}, F_{1P})$ space is shown in Fig.~\ref{fig:mf1p_mag}, based on Table~2 of \citet{mil20} and Table~3 of \citet{don21}.  Light-blue pentagons and deep-blue diamonds depict LMC and SMC clusters, respectively.   They occupy the outer-halo region of the diagram, in agreement with their less advanced dissolution in the gentle tidal field of the Magellanic Clouds, combined to their intermediate age (from 2\,Gyr for NGC1978 to 10.5\,Gyr for NGC121).  Note that when comparing their location to the intial track $F_{1P}(m_{init})$, we implicitly assume that Galactic \gcs and Magellanic Clouds clusters share the same $F_{1P}(m_{init})$ relation, which may or may not be true.  We also stress that the red tracks and the location of the LMC and SMC clusters cannot be compared on a one-by-one basis given their age difference (12\,Gyr for the tracks and less than 10.5\,Gyr for the LMC and SMC clusters).      

Finally, we also add the 4 metal-poor \gcs of the Fornax dwarf spheroidal galaxy, Fornax~1, Fornax~2, Fornax~3 and Fornax~5, with the pristine-star fractions from \citet{lar14}.  Cluster masses have been inferred from the visual absolute magnitudes given in Table~1 of \citet{lar12} with an assumed mass-to-light ratio of 2.0.  Fornax \gcs are depicted in Fig.~\ref{fig:mf1p_mag} as purple triangles.   Fornax~2 and 3 (the two most upper-right points) are located among outer halo globular clusters, but to the left of most Magellanic Clouds clusters, as expected given their older ages \citep{zin81}.  The location of Fornax~1 (the least massive) and Fornax~5 is less consistent with them having evolved in a weak tidal field, although the size of the error bars does not allow one to draw definitive conclusions.  We stress again that whether multiple-populations clusters form in a similar way in large and dwarf galaxies remains to be seen. 

\section{The mass fraction of 2P stars \\ among Halo Field Stars}
\label{sec:2PHFS}
In the Galactic halo, the observed fraction of field stars bearing the chemical signature of hot hydrogen burning  is reported to be of order a few per cent.  \citet{mar11} infer a fraction of 3\,\% of CN-strong stars (i.e. N-enhanced stars) in a sample of 561 metal-poor red giants.  \citet{car10a} obtain an even lower fraction (1.4\,\%) of Na-rich stars among field metal-poor stars.  Because of our assumption that each cluster loses its 1P- and 2P-stars with the same likelihood, our model may seem at odds with these observations.  Yet, it does  match the observed low fraction of 2P-/CN-strong field stars in the Galactic halo, for two reasons.  Firstly, 2P-star formation is restricted to a limited fraction of the cluster mass spectrum, that is, in clusters initially more massive than $m_{init}=F_{bound}^{VR} m_{th} = 4 \cdot 10^5\,\Ms$ (Eq.~\ref{eq:f1p-init}).  Secondly, the high mass of these clusters makes many of them resilient to tidal stripping (at least for a canonical IMF), thereby helping them retain their stars in general, and their 2P-stars in particular.  In other model classes \citep{dec07,derc08}, the formation of 2P stars is restricted to cluster inner regions, and this is their central location inside clusters that accounts for their small fraction in the Galactic field.  In contrast, in the present model, the small fraction of 2P- stars shed into the field stems from their forming in massive - hence resilient - clusters.         

In what follows, we use upper case 'M' to denote cluster masses integrated over all, or part of, the cluster mass spectrum and lower case 'm' to denote individual cluster masses (e.g. $M_{ecl}$ is the total stellar mass formed in gas-embedded clusters, while $m_{ecl}$ is the mass of an embedded cluster).  

We assume that the stellar halo has been built from two main sources: (1)~a population of gas-embedded clusters of total stellar mass $M_{ecl}$ and containing all stars formed in-situ, and (2)~the accretion of dwarf galaxies of total present-day stellar mass $M_{acc}$.  This yields   
\begin{equation}
M_{halo}=F_{bound}^{StEv}M_{ecl}+M_{acc}\,,
\end{equation}
with the present-day stellar mass of the halo being $M_{halo} \simeq 10^9\,\Ms$ \citep{sun91}.  

We assume that all embedded clusters have a canonical stellar IMF (i.e. $\alpha_3=2.3$, hence $F_{bound}^{StEv} \simeq 0.6{\rm -}0.7$), and that the cluster mass distribution obeys a single-index power law
\begin{equation}
{\rm d}N_{ecl} = k_M m_{ecl}^{-\alpha} {\rm d}m_{ecl} 
\label{eq:eclmf}
\end{equation}
with lower and upper limits $m_{low}$ and $m_{up}$, respectively.  A single-index power-law mass function with $\alpha=2$ is observed for the young compact and massive clusters formed in starbursts and galaxy mergers \citep[e.g.][]{bik03,fal09}.  Whether the initial globular cluster mass function is a similar power-law remains much debated, however \citep[][see also Sec.~\ref{sec:conc}]{ves98,fz01,par07}.  For the sake of simplicity, we assume a power-law initial globular cluster mass function with $\alpha=2$ as it allows us to derive analytical expressions of  straithgforward physical interpretation.  More complex initial cluster mass functions will be considered in future papers.  We adopt $m_{low}=10^2\,\Ms$, $m_{up}$ being allowed to vary.  The normalization factor $k_M$ obeys:
\begin{equation}
k_M = \frac{M_{ecl}}{ln\left(\frac{m_{up}}{m_{low}}\right)}\;. 
\end{equation}   

We now aim at constraining the mass fraction of 2P stars (1)~in the system of gas-embedded clusters, equivalently among all in-situ stars, (2)~among halo field stars ('HFS') after cluster violent relaxation, and (3)~among halo field stars at an age of about 12\,Gyr, when dwarf galaxies have been accreted into the Galactic halo.

\subsection{The fraction of 2P-stars in the gas-embedded cluster phase ($F_{2P,ecl}$)}
\label{ssec:2PHFS_ecl}
For a power-law cluster mass function with $\alpha=2$ (Eq.~\ref{eq:eclmf}), the ratio between the total mass in 2P stars, $M_{ecl,2P}$, and the total stellar mass $M_{ecl}$, in gas-embedded clusters is
\begin{equation}
\begin{split}
F_{2P,ecl} = & \frac{M_{ecl,2P}}{M_{ecl}} = \frac{ln\left(\frac{m_{up}}{m_{th}}\right)-\left(1-\frac{m_{th}}{m_{up}}\right)}{ln\left(\frac{m_{up}}{m_{low}}\right)} \\
           = & \frac{M_{ecl}-M_{ecl,1P}}{M_{ecl}} = 1 - F_{1P,ecl}\;.
\end{split}
\label{eq:f2p-ecl}
\end{equation}
Note that $F_{1P,ecl}$ denotes the mass fraction of 1P stars for all gas-embedded clusters, while $F_{1P}(m_{ecl})$ is the mass fraction of 1P-stars in an embedded cluster of mass $m_{ecl}$ (Eq.~\ref{eq:f1p-ecl}).  
For $m_{low}=10^2\,\Ms$, $m_{th}=10^6\,\Ms$ and $m_{up}=5\cdot10^6\,\Ms$, $F_{2P,ecl}=7.5\,\%$.  For $m_{up}=10^7\,\Ms$, $F_{2P,ecl}=12.2\,\%$.  We discuss the sensitivity of $F_{2P,ecl}$ to the cluster mass spectrum upper limit $m_{up}$ later in this section.  Note that, at this stage, there is no need to define the fraction of 2P-stars in the halo field since halo field stars do not exist yet (gas-embedded clusters have not violently relaxed yet, and dwarf galaxies have not been accreted yet).  \\

\subsection{The fraction of 2P-stars among halo field stars after violent relaxation ($F_{2P,HFS}^{VR}$)} 
\label{ssec:2PHFS_vr}
Following their formation inside gas clumps, clusters expel their gas, expand and lose a fraction $(1-F_{bound}^{VR})$ of their stars (Sec.~\ref{ssec:vr}).  What is the fraction of 2P stars in the field at the end of violent relaxation, $F_{2P,HFS}^{VR}$?  
Here, the reader may recall two model assumptions: (i)~the fraction $F_{bound}^{VR}$ of stars that clusters retain by the end of violent relaxation is independent of their mass $m_{ecl}$, and (ii)~each multiple-populations cluster loses its 1P- and 2P-stars with the same probability.
Therefore, the masses in 1P- and 2P-stars in clusters that have returned to virial equilibrium after gas expulsion are
\begin{subequations}
\begin{equation}
m_{init,1P} = F_{bound}^{VR} m_{th} 
\end{equation} 
\begin{equation}
m_{init,2P} = F_{bound}^{VR} m_{ecl,2P}
\end{equation} 
\end{subequations}
with 
\begin{equation}
\begin{split}
m_{ecl} = & m_{th} + m_{ecl,2P} \\
        = & m_{th} + [1-F_{1P}(m_{ecl})] \times m_{ecl}
\end{split} 
\end{equation}
and
\begin{equation}
m_{init} = m_{init,1P} + m_{init,2P}\;. 
\end{equation}

As a result of the two assumptions above, the total fractions of $2P$-stars inside clusters and in the field  after violent relaxation, $F_{2P,init}$ and $F_{2P,HFS}^{VR}$, respectively, equate with the 2P-star fraction in the embedded-cluster phase (Eq.~\ref{eq:f2p-ecl}):
\begin{equation}
F_{2P,HFS}^{VR} = F_{2P,init} = F_{2P,ecl} = \frac{ln\left(\frac{m_{up}}{m_{th}}\right)-\left(1-\frac{m_{th}}{m_{up}}\right)}{ln\left(\frac{m_{up}}{m_{low}}\right)}\;. 
\label{eq:2PHFS_vr}
\end{equation}
We have seen in Sec.~\ref{ssec:2PHFS_ecl} that for an embedded-cluster mass upper limit  $m_{up}=10^7\,\Ms$, $F_{2P,ecl}=0.12$, hence $F_{2P,HFS}^{VR}=0.12$.  This is significantly higher than its observed present-day counterpart \citep[0.03,][]{mar11}.  However, after violent relaxation, the fraction of 2P-stars in the halo field is bound to decrease: low-mass clusters, those made of 1P-stars only, are those most disrupted by the long-term secular evolution, and 2P field stars do not seem to be present in dwarf galaxies \citep{nor17,sal19} 

\subsection{The present-day fraction of 2P-stars among halo field stars ($F_{2P,HFS}^{12\,Gyr}$)} 
\label{ssec:2PHFS_12G}
To infer the mass fraction of 2P-stars in the halo field at an age of $\simeq 12$\,Gyr, $F_{2P,HFS}^{12\,Gyr}$, is obviously a complex task.  Not only does it require a cluster dissolution model, it also calls for various  assumptions such as the initial cluster mass function, the cluster stellar IMF, the initial cluster radial distribution across the Galactic halo, and their orbit distribution.  Last but not least, the contribution of accreted dwarf galaxies to the halo needs to be accounted for.  Although such a task remains beyond the scope of this paper, it is doable to constrain $F_{2P,HFS}^{12\,Gyr}$ by considering two extreme scenarios, one that maximizes the fraction of 2P-stars in the halo field, and one that minimizes it.  We now derive these upper and lower limits, $\left. F_{2P,HFS}^{12\,Gyr}\right]^+$ and $\left. F_{2P,HFS}^{12\,Gyr}\right]_-$, respectively. \\ 

{\it Upper limit $\left. F_{2P,HFS}^{12\,Gyr}\right]^+$:} An upper limit on the current mass fraction of 2P-stars in the halo field occurs if all massive clusters - those having formed 2P-stars - dissolve.  This assumption necessarily implies that {\it all} clusters, from the low-mass to the high-mass ones, dissolve, and that the stellar halo is made of field stars only.  This is not such an extreme assumption since \gcs today represent a minor fraction of the stellar halo mass, namely, $\simeq 2$\,\% \citep{sun91,mor93}.  The total mass in 2P-stars in the halo is then the total mass in 2P-stars in embedded clusters, corrected for stellar evolutionary mass-losses, $F_{bound}^{StEv}M_{ecl,2P}$, and augmented by the mass of those accreted, $M_{acc,2P}$.  Our upper limit is therefore:
\begin{equation}
\left. F_{2P,HFS}^{12\,Gyr}\right]^+ = 
\frac{ F_{bound}^{StEv} M_{ecl,2P} + M_{acc,2P}}{ F_{bound}^{StEv} M_{ecl} + M_{acc}}\;.
\label{eq:2PHFS_all1}
\end{equation}     
Writing the accreted mass as a function of the in-situ mass, $M_{acc}=f F_{bound}^{StEv} M_{ecl}$, Equation~\ref{eq:2PHFS_all1} becomes:
\begin{equation}
\left. F_{2P,HFS}^{12\,Gyr}\right]^+  = \frac{F_{2P,ecl}+fF_{2P,acc}}{1+f}\,,
\label{eq:2PHFS_all2}
\end{equation} 
with $F_{2P,acc}=M_{acc,2P}/M_{acc}$, the mass fraction of 2P stars in the accreted component (clusters and field stars alike).  We now need an estimate of $F_{2P,acc}$. 
 
Nitrogen- and Na-rich stars have so far remained undetected in samples of dwarf galaxy field stars \citep[e.g.][]{nor17,sal19}.    
Dwarf galaxies, however, host globular clusters, and nuclear \gcs \citep[e.g. NGC6715 for the Sagittarius galaxy,][]{car10b}, whose accretion and evaporation contribute to the Galactic 2P field star population. Nuclear \gcs provide about $1-2$\,\% of the total luminosity of their host dwarf galaxy \citep{fbh02,geo09}, and we take this luminosity ratio as a proxy to the corresponding mass  ratio.  To assess the contribution of non-nuclear globular clusters, we consider the Fornax dwarf spheroidal galaxy, one of the highest specific-frequency galaxies \citep[see the Appendix in ][]{az98}.  Its 5 globular clusters represent $\simeq 2$\,\% of the total stellar mass it has formed\footnote{The stellar mass formed by Fornax is about $4.3 \cdot 10^7\,\Ms$ \citep{deB12}.  Assuming a mass-to-light ratio of 2.0, the total mass of its 5 \gcs is $\simeq 8 \cdot 10^5\,\Ms$ \citep[based on the cluster absolute visual magnitudes listed in Table~1 of][]{lar12}.  }.  All together, nuclear and non-nuclear \gcs in dwarf galaxies thus represent, at most, a fraction $f_{GC,acc} \simeq 4$\,\% of their host galaxy stellar mass.     

Since we are after an upper limit of $F_{2P,HFS}^{12\,Gyr}$, we assume that all accreted \gcs are made of 2P- stars only.  This is an excellent assumption for dwarf galaxy nuclei since the fraction of 2P-stars in $\omega$~Cen is about 90\,\% \citep[][]{mil17}.  Therefore, an upper limit on the 2P-star fraction integrated over a sample of dwarf galaxies is 
\begin{equation}
\left. F_{2P,acc}\right]^+ \simeq 4\,\%\;.
\label{eq:4pc}
\end{equation}  
  
This result alone is remarkably close to the observational results of \citet[][1.4\,\%]{car10a} and \citet[][3\,\%]{mar11}.  That is, a Galactic halo made entirely of fully dissolved dwarf galaxies and of their dissolved \gcs has a fraction of 2P field stars no larger than a few per cent, as observed.  However, what if in-situ stars contribute as much as 50\,\% of the halo stellar mass?  The accreted and in-situ masses are then comparable, $f\simeq 1$ and $M_{acc} = F_{bound}^{StEv} M_{ecl} = M_{halo}/2 \simeq 5 \cdot 10^8\,\Ms$ \citep[for $M_{halo}\simeq 10^9\,\Ms$,][]{sun91}, and Eq.~\ref{eq:2PHFS_all2} becomes
\begin{equation}
\left. F_{2P,HFS}^{12\,Gyr}\right]^+ = \frac{F_{2P,ecl}+\left. F_{2P,acc}\right]^+}{2}\;.
\label{eq:2PHFS_all3}
\end{equation}  
Equation~\ref{eq:2PHFS_all3} defines an upper limit on the present-day mass fraction of 2P stars in the halo field because of the following three assumptions: \\
-1- all massive clusters dissolve, and their content remains in the halo (dynamical friction is not - yet - taken into account); \\
-2- accreted \gcs and nuclear \gcs are assumed to be made of 2P stars only; \\
-3- accreted dwarf galaxies are assumed to be high specific-frequency galaxies, thereby maximizing their cluster mass fraction. 

Extending the example introduced in Sec.~\ref{ssec:2PHFS_ecl}, for which we find $F_{2P,ecl}=7.5\,\%$ with $m_{low}=10^2\,\Ms$, $m_{th}=10^6\,\Ms$ and $m_{up}=5\cdot10^6\,\Ms$, we obtain an upper limit on the mass fraction of 2P-stars in the halo field of $\left. F_{2P,HFS}^{12\,Gyr}\right]^+ = 5.7\,\%$.

What is the corresponding fraction $F_{HFS,Halo}^{12Gyr}$ of field stars in the Galactic halo?  That is, what is the fraction of stars that formed either in in-situ \gcs or in accreted dwarf galaxies, and that today populate the halo field?  
Since all clusters -- in-situ or accreted -- are assumed to dissolve, our upper limit $\left. F_{2P,HFS}^{12\,Gyr}\right]^+$ couples with a fraction of field stars in the Galactic halo of
\begin{equation}
F_{HFS,Halo}^{12Gyr} = 1.00\;.
\label{eq:HFS_all}
\end{equation}

We now move to estimating a lower limit on $F_{2P,HFS}^{12\,Gyr}$.\\

{\it Lower limit $\left. F_{2P,HFS}^{12\,Gyr}\right]_-$:} The fraction of 2P-stars in the halo field reaches a minimum if, during secular evolution: (1)~all single-population clusters dissolve, while the more massive multiple-populations clusters survive unscathed, and (2) accreted dwarf galaxies add only 1P-stars to the halo field (i.e. accreted \gcs do not dissolve at all).  This ensures that, at an age of 12\,Gyr, the only 2P field stars are those that were released by violently relaxing in-situ clusters.  

The total mass in single-population clusters (i.e. clusters for which $m_{ecl} < m_{th}$) is, after violent relaxation:
\begin{equation}
M_{Sngl-Pop-Cl}^{VR} = F_{bound}^{VR} M_{ecl} \frac{ln\left(\frac{m_{th}}{m_{low}}\right)}{ln\left(\frac{m_{up}}{m_{low}}\right)}\;.
\end{equation} 
Their dissolution by an age of 12\,Gyr adds to the halo field, but not to the 2P field-star population.  The total mass in halo field stars at an age of 12\,Gyr then includes three components:
\begin{equation}
\begin{split}
M_{HFS}^{12Gyr} = & F_{bound}^{StEv}(1-F_{bound}^{VR}) M_{ecl} \\
        + & F_{bound}^{StEv} M_{Sngl-Pop-Cl}^{VR} \\
        + & (1-f_{GC,acc}) M_{acc}\,,
\end{split}
\label{eq:Mhfs}
\end{equation}
namely, the stars removed from all in-situ clusters during violent relaxation (first term), the stars arising from the (complete) secular dissolution of all single-population clusters (second term), and the accreted field stars (third term).  For the parameter values previously used ($m_{low}=10^2\,\Ms$, $m_{th}=10^6\,\Ms$, $m_{up}=5\cdot10^6\,\Ms$, $F_{bound}^{VR}=0.4$, $F_{bound}^{StEv} = 0.7$, $M_{ecl} = 7 \cdot 10^8\,\Ms$, $f_{GC,acc}=0.04$, and $M_{acc}=5 \cdot 10^8\,\Ms$), we find $M_{HFS}^{12Gyr}=9.5 \cdot 10^8\,\Ms$.

The lower limit on the 2P-star fraction in the halo field is then 
\begin{equation}
\begin{split}
\left. F_{2P,HFS}^{12\,Gyr}\right]_- = & \frac{F_{bound}^{StEv}(1-F_{bound}^{VR}) M_{ecl,2P} }{M_{HFS}^{12Gyr}} \\
                                  = &  \frac{F_{bound}^{StEv}(1-F_{bound}^{VR}) F_{2P,ecl} M_{ecl} }{M_{HFS}^{12Gyr}}  
\end{split}
\label{eq:2PHFS_spd}
\end{equation}
where the numerator represents the total mass in 2P stars shed into the field by the end of violent relaxation, corrected for stellar evolution.  

The corresponding fraction of field stars in the Galactic halo obeys:
\begin{equation}
F_{HFS,Halo}^{12Gyr} \\ = \frac{M_{HFS}^{12Gyr}}{F_{bound}^{StEv}M_{ecl}+M_{acc} }
\label{eq:HFS_spd}
\end{equation}

Using the same parameter values as for Eq.~\ref{eq:Mhfs}, the lower limit on the mass fraction of 2P-stars in the halo field is $\left. F_{2P,HFS}^{12\,Gyr}\right]_- = 2.3\,\%$, and field stars represent $M_{HFS}^{12Gyr}/M_{Halo}=95\,\%$ of the stellar halo mass. 

\begin{figure}
\begin{center}
\includegraphics[width=0.49\textwidth]{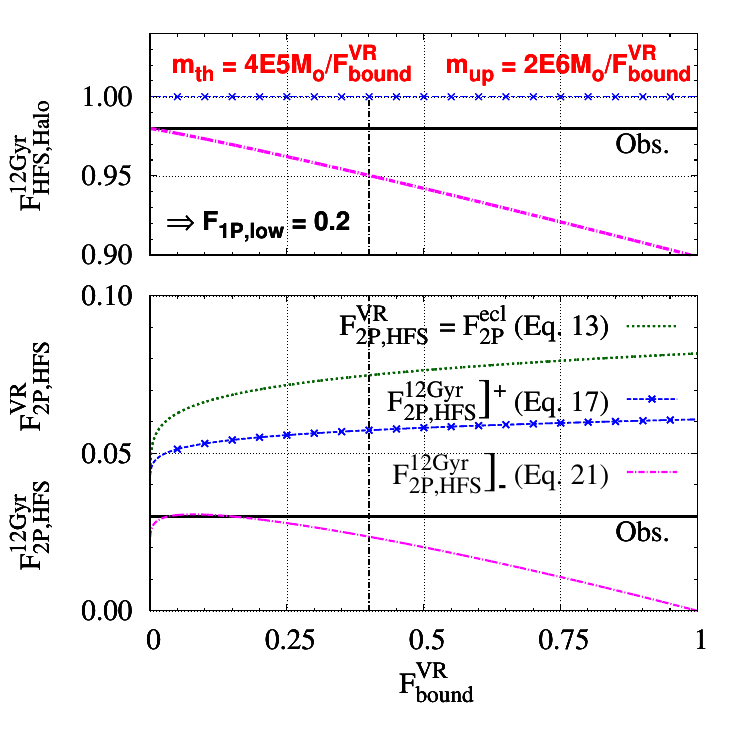}
\caption{Top panel: Present-day fraction of field stars in the Galactic halo for the upper and lower limits on the 2P field star fraction $F_{2P,HFS}^{12Gyr}$ shown in the bottom panel (blue line with asterisks and dash-dotted magenta line, respectively).  Bottom panel: Fraction of 2P-stars in the halo field.  The dotted green line is the result after violent relaxation (i.e.~at  secular evolution onset), while the blue line with asterisks and the magenta dash-dotted line are the upper and lower limits at an age of 12\,Gyr.  In each panel, the solid horizontal line is the observational constraint \citep[][respectively]{sun91,mar11}.  The dash-dotted vertical lines mark the bound fraction used in Fig.~\ref{fig:2PHFS_mup}.  Note the different $y$-axis ranges.  $F_{1P,low}$ is the lowest 1P-star fraction achieved by an embedded cluster (i.e. $F_{1P,low}=m_{th}/m_{up}$; Eq.~\ref{eq:f1p-ecl})}
\label{fig:2PHFS_Fb}
\end{center} 
\end{figure}

Equations \ref{eq:2PHFS_all3}, \ref{eq:2PHFS_spd} and \ref{eq:HFS_spd} are discussed in greater depth in Fig.~\ref{fig:2PHFS_Fb} and Sec.~\ref{ssec:2PHFS_disc}.

\subsection{Meeting the observational constraints} 
\label{ssec:2PHFS_disc}
Results from Sec.~\ref{ssec:2PHFS_vr} and \ref{ssec:2PHFS_12G} are shown in Fig.~\ref{fig:2PHFS_Fb} as a function of the bound fraction $F_{bound}^{VR}$ at the end of violent relaxation.  The top panel depicts the mass fraction of field stars in the halo $F_{HFS,Halo}^{12\,Gyr}$ (Eqs~\ref{eq:HFS_all} and \ref{eq:HFS_spd}).  The bottom panel presents the mass fraction of 2P stars in the halo field after violent relaxation $F_{2P,HFS}^{VR}$ (Eq.~\ref{eq:2PHFS_vr}), as well as the upper and lower limits at an age of 12\,Gyr (Eqs~\ref{eq:2PHFS_all3} and \ref{eq:2PHFS_spd}).  As previously (Sec.~\ref{sec:comp}), we tie the bound fraction and the embedded-cluster mass threshold such that $F_{bound}^{VR} m_{th} = 4\cdot 10^5\,\Ms$.  We also tie the embedded-cluster mass upper limit $m_{up}$ and the bound fraction $F_{bound}^{VR}$ such that $F_{bound}^{VR} m_{up} = 2\cdot 10^6\,\Ms$.  That is, the most massive cluster at secular evolution onset has a mass of $2\cdot 10^6\,\Ms$.  This ensures that the lowest $F_{1P}$ value is $F_{1P,low} \simeq m_{th} / m_{up} = 4\cdot 10^5\,\Ms / 2\cdot 10^6\,\Ms \simeq 0.2$, as observed \citep[$F_{1P}=0.18$ for NGC104 and NGC6205,][]{don21}\footnote{Although the 1P-star fraction of $\omega$~Cen is even lower \citep[$F_{1P}=0.09$][]{mil17}, we do not consider it here as $\omega$~Cen is the tidally-stripped nucleus of a dwarf galaxy, rather than a genuine globular cluster \citep[e.g.][]{hil00,car10b}.}.  The impact of varying $m_{up}$ is addressed in Fig.~\ref{fig:2PHFS_mup}.  As previously, the lower limit on the embedded cluster mass spectrum is $m_{low}=100\,\Ms$.  
In each panel of Fig.~\ref{fig:2PHFS_Fb}, the solid horizontal black line is the observational result, namely, $F_{HFS,Halo}^{obs}=0.98$ in top panel \citep{sun91} and $F_{2P,HFS}^{obs}=0.03$ in bottom panel \citep{mar11}.  

The dissolution of all clusters, including the high-mass ones, by an age of 12\,Gyr maximises both the field star fraction in the halo and the 2P-star fraction in the field (blue lines with asterisks).  In the bottom panel, the blue line corresponds to the mean of the green curve and of $f_{GC,acc}=\left. F_{2P,acc}^{12\,Gyr}\right]^+=0.04$ (see Eq.~\ref{eq:2PHFS_all3}).  Note that Eq.~\ref{eq:2PHFS_all3} defines a firm upper limit since the contribution of 2P stars from the accreted component has been maximized as well (that is, to obtain Eq.~\ref{eq:4pc}, accreted galaxies of high specific-frequency, and accreted \gcs made of 2P stars only, have been assumed).

Conversely, when secular evolution dissolves only single-population clusters, while sparing the high-mass multiple-populations ones, the only source of 2P stars is violent relaxation.  The 2P-star fraction is therefore minimum, and an increasing bound fraction $F_{bound}^{VR}$ decreases both the 2P-star fraction in the field $F_{2P,HFS}^{12Gyr}$ and the field star fraction in the halo $F_{HFS,Halo}^{12Gyr}$ (dash-dotted magenta lines in both panels).  When $F_{bound}^{VR}=1$, 2P-stars are shed in the field neither through violent relaxation ($F_{bound}^{VR}=1$), nor through secular evolution (only single-population clusters dissolve).  The 2P-star fraction is thus zero.  When $F_{bound}^{VR}=0.0$, all in-situ/embedded clusters release their 2P stars straight after gas expulsion, and the field 2P-star content is later on diluted by a factor $\simeq 2$ as accreted galaxies add their 1P-field stars to the halo (assuming comparable contributions of the in-situ and accreted components to the Galactic halo).  That is, when $F_{bound}^{VR} \gtrsim 0.0$, the magenta curve is half of the green curve.
 
Figure~\ref{fig:2PHFS_Fb} shows that the lower and upper limits on $F_{HFS,Halo}^{12Gyr}$ and $F_{2P,HFS}^{12Gyr}$ bracket their respective observational counterparts.  The present model is therefore able to yield a low 2P-star fraction in the halo field, as observed, in spite of 1P and 2P stars being equally-likely lost from clusters.  This stems from the high cluster mass threshold ($m_{ecl} > m_{th}=10^6\,\Ms$) imposed for embedded clusters to form 2P stars.  

However, the reader could rightly comment that raising the upper limit of the cluster mass spectrum would also raise all curves in the bottom panel of Fig.~\ref{fig:2PHFS_Fb}.  This would actually equate with adding major 2P-star providers to the cluster system.  So far, none of the embedded clusters has a 2P star fraction exceeding 80\,\% (i.e. $F_{1P,low} \simeq F_{bound}^{VR} m_{th} / ( F_{bound}^{VR} m_{up} ) \simeq 0.2$).  This raises two questions: (1)~What if the initial cluster population includes clusters with $F_{1P}<<0.2$?  (2)~Why do we not observe such (genuine) globular clusters?

\subsection{Impact of the embedded-cluster mass upper limit: meeting dynamical friction (again)} 
\label{ssec:2PHFS_mup}

\begin{figure}
\begin{center}
\includegraphics[width=0.49\textwidth]{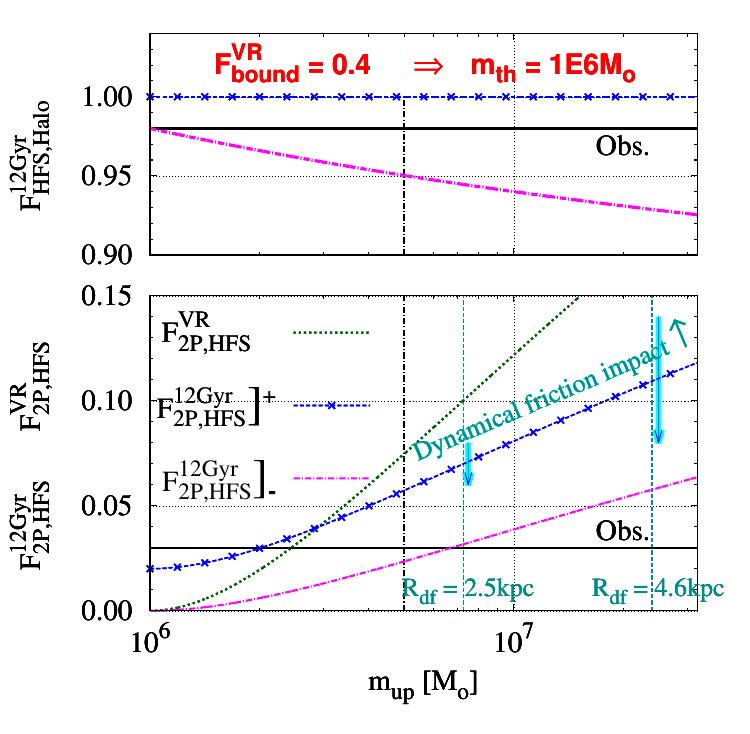}
\caption{Same as Fig.~\ref{fig:2PHFS_Fb}, but the field star fraction in the halo (top panel) and the 2P-star fraction in the halo field (bottom panel) are depicted in dependence of the embedded-cluster mass upper limit $m_{up}$, rather than as a function of the bound fraction $F_{bound}^{VR}$ at the end of violent relaxation.  The dark-cyan dashed vertical lines mark the masses corresponding to the dynamical friction limits $R_{df}=2.5$\,kpc and $R_{df}=4.6$\,kpc.  The downwards blue/cyan arrows indicate that the upper limit $\left. F_{2P,HFS}^{12\,Gyr}\right]^+$ is expected to decrease once dynamical friction is accounted for.  }
\label{fig:2PHFS_mup}
\end{center} 
\end{figure} 

Figure~\ref{fig:2PHFS_mup} depicts the same equations (with identical color/line coding) as Fig.~\ref{fig:2PHFS_Fb}, but as a function of $m_{up}$, rather than as a function of $F_{bound}^{VR}$.  A bound fraction $F_{bound}^{VR}=0.4$ is imposed, which yields $m_{th}=10^6\,\Ms$.  The cluster mass lower limit $m_{low}=10^2\,\Ms$ is unchanged.  The vertical dash-dotted black lines in Figs~\ref{fig:2PHFS_Fb} and \ref{fig:2PHFS_mup} mark the parameter set common to both figures, i.e. $F_{bound}^{VR}=0.4$, $m_{th}=10^6\,\Ms$ and $m_{up}=5\cdot 10^6\,\Ms$.  That is, the dash-dotted vertical lines render identical fractions in both figures (e.g. $F_{2P,ecl}=7.5$\,\%).  

By increasing the embedded-cluster mass upper limit $m_{up}$, clusters containing larger fractions of 2P stars get added  to the cluster system (Eq.~\ref{eq:f1p-ecl}).  Integrated 2P-star fractions rise with $m_{up}$ as a result (bottom panel in Fig.~\ref{fig:2PHFS_mup}).  When $m_{up}\simeq 7\cdot 10^6\,\Ms$, the lower limit on $F_{2P,HFS}^{12Gyr}$ (dash-dotted magenta line) becomes larger than the observational constraint.  Could the embedded-cluster mass upper limit $m_{up}$ be around $7 \cdot 10^6\,\Ms$?  Is the number of observed clusters with $F_{1P}\lesssim 0.2$ consistent with $m_{up} \lesssim 7 \cdot 10^6\,\Ms$?

With the adopted model parameters ($M_{ecl}=7.8 \cdot 10^8\,\Ms$, $m_{low}=10^2\,\Ms$, $m_{up}=7\cdot 10^6\,\Ms$ and $\alpha =2$), we predict 3-4 clusters with a mass $m_{ecl}$ between $5 \cdot 10^6\,\Ms$ and $7 \cdot 10^6\,\Ms$, thus with $0.14 \leq F_{1P} \leq 0.2$.  That is, at the time of its formation, we predict the in-situ halo \gc system to contain 3-4 clusters with a 1P-star fraction less than 20\,\%.  Yet, we should observe fewer of these clusters since, in the mean time, dynamical friction has removed those closest to the Galactic center (see Fig.~\ref{fig:fred_fric}).  The dynamical friction limit for an embedded-cluster mass $m_{ecl} \simeq 6 \cdot 10^6\,\Ms$ is $R_{df} \simeq 1.44 \sqrt{0.4 \times 6} \simeq 2.2\,$kpc (Sec.~\ref{ssec:dynfric}), where the factor $0.4$ is the violent relaxation bound fraction adopted in Fig.~\ref{fig:2PHFS_mup}.  In other words, clusters born with an embedded mass $m_{ecl} \simeq 6 \cdot 10^6\,\Ms$ (hence $F_{1P} =0.17$) and located within 2.2\,kpc from the Galactic center have been engulfed by the bulge, and can no longer be observed.  What fraction of the clusters do they represent?  

The Galactic halo globular cluster system is centrally-concentrated, its number density $n_{GCS}$ plummeting with Galactocentric distance as $D_{gal}^{-3.5}$, i.e.: 
\begin{equation}
n_{GCS}(D_{Gal}) \propto \left( 1 + \frac{D_{gal}}{D_c} \right)^{-\gamma}
\label{eq:rhoGCS}
\end{equation}
where $\gamma = 3.5$ and $D_c=1$\,kpc (e.g. \citealt{djm94}; see also fig.~2 in \citealt{mcl99}).  
Assuming a halo extending from 1.0 to 120.0\,kpc, we infer that 13\,\% of the clusters are located within the dynamical friction limit $R_{df}=2.2$\,kpc.  Initially, however, the \gc system is more centrally-concentrated than it is today, because secular evaporation is more damaging at shorter Galactocentric distances.  If $\gamma = 4.5$ in Eq.~\ref{eq:rhoGCS}, the fraction of clusters with an embedded mass $m_{ecl} \simeq 6 \cdot 10^6\,\Ms$ and eventually removed by dynamical friction rises to about one-third.  We therefore expect to observe 2-3 clusters with $0.14 \leq F_{1P} \leq 0.2$, out of the 3-4 that initially formed, in agreement with the two such clusters actually observed (i.e. NGC104 and NGC6205, for which $F_{1P}=0.18$).  

Therefore, with the adopted model parameters, an embedded-cluster mass upper limit $m_{up} \simeq 7 \cdot 10^6 \,\Ms$ is  consistent with (1)~the lowest $F_{1P}$ values observed ($F_{1P}=0.18$), (2)~the number of observed \gcs showing these $F_{1P}$ values (2 clusters), and (3)~the observed fraction of 2P stars in the halo field ($F_{2P,HFS}^{obs}=0.03$).  Here we remind the reader of the key model assumptions: a power-law mass function of slope $-\alpha=-2$ for the embedded clusters, an in-situ halo mass $M_{ecl} = 7.1 \cdot 10^8\,\Ms$ (reduced to a present-day mass of $5 \cdot 10^8\,\Ms$ by stellar-evolution mass losses), and Eq.~\ref{eq:f1p-ecl} to relate the embedded mass of clusters to their 1P-/2P-star fractions.  We also recall that the 1P-/2P-star fractions inside clusters are assumed to stay constant as clusters evolve.

By how much does dynamical friction affect the curves in Fig.~\ref{fig:2PHFS_mup}?  Dynamical friction alters neither the 2P-star fraction after violent relaxation $F_{2P,HFS}^{VR}$ (time-scale too short by 2-3 orders of magnitude), nor the lower limit $\left. F_{2P,HFS}^{12\,Gyr}\right]_-$ at an age of 12\,Gyr (see Eqs~\ref{eq:Mhfs} and \ref{eq:2PHFS_spd}).  In contrast, the upper limit $\left. F_{2P,HFS}^{12\,Gyr}\right]^+$ is to be reduced, the larger $m_{up}$, the greater the dynamical friction impact, the stronger the reduction of $\left. F_{2P,HFS}^{12\,Gyr}\right]^+$.  To quantify the latter, we assume that dynamical friction affects only in-situ clusters.  We divide the halo in nested spherical shells and, for each shell (i.e. for each Galactocentric distance), we obtain the embedded-cluster mass $m_{df}$ beyond which clusters get destroyed by dynamical friction within 12\,Gyr. 
We then calculate, for each shell, the total masses in stars and 2P-stars that clusters in the mass range $m_{df}$-$m_{up}$ contain, and we remove them from the halo mass budget.  The initial radial density profile of the halo obeys Eq.~\ref{eq:rhoGCS}, with either $\gamma=3.5$ (i.e. the initial and present-day slopes of the halo density profile are similar\footnote{The density profile fitting the metal-poor ([Fe/H]$<-1$) \gc system in fig.~2 of \citet{mcl99} provides results indistinguishable from $\gamma=3.5$}) or $\gamma=4.5$ (i.e. the halo initial density profile is steeper than it is today).       

\begin{figure}
\begin{center}
\includegraphics[width=0.49\textwidth]{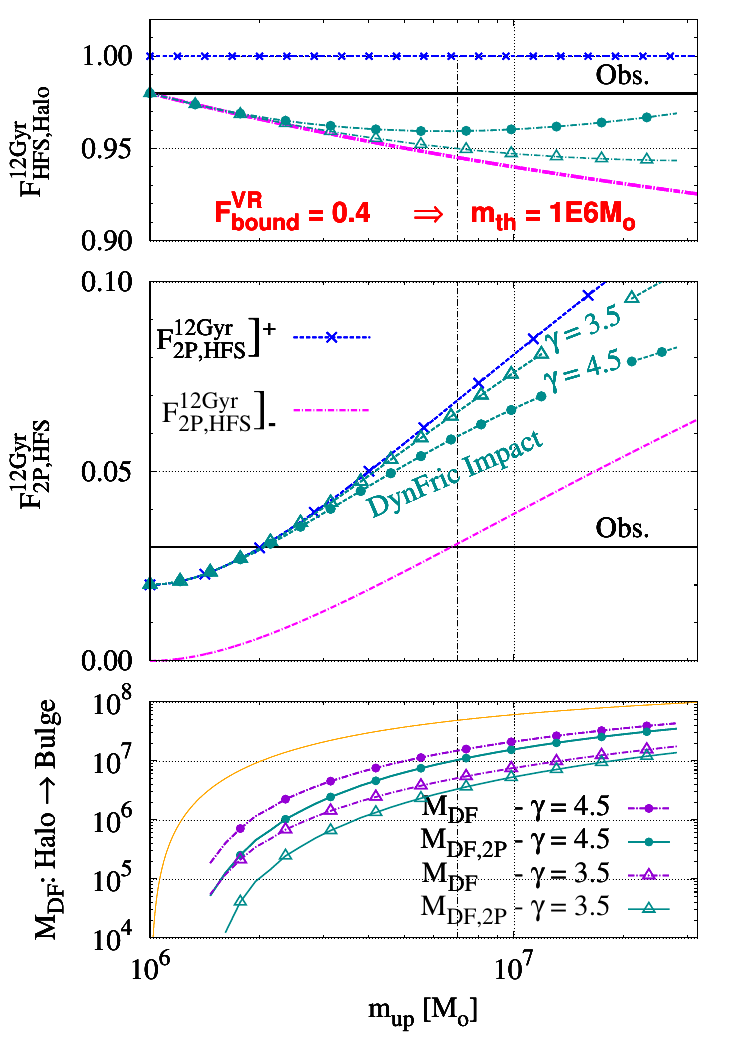}
\caption{
Top panel: fraction of field stars in the Galactic halo for two cases: secular dissolution of all clusters (blue line with asterisks), corresponding to the upper limit  $\left. F_{2P,HFS}^{12Gyr}\right]^+$, and secular dissolution of the single-population clusters only (dash-dotted magenta line), corresponding to the lower limit $\left. F_{2P,HFS}^{12Gyr} \right]_-$. In the second case, the dark-cyan lines show the impact of dynamical friction, which decreases the halo mass by removing  massive clusters close to the Galactic center.  Two halo initial density profiles (Eq.~\ref{eq:rhoGCS} with $D_c=1$\,kpc) are tested: $\gamma=3.5$ (open triangles) and $\gamma=4.5$ (plain circles).  The dash-dotted vertical line mark our estimate of the embedded-cluster mass upper limit ($m_{up} \simeq 7\cdot 10^6\,\Ms$).    
Middle panel: Upper and lower limits on the 2P-star fraction in the halo field at an age of 12\,Gyr (same color-coding as in Fig.~\ref{fig:2PHFS_mup}) with, for the upper limit, the  corrections due to dynamical friction.  
Bottom panel: Mass in stars ($M_{DF}$) and 2P-stars ($M_{DF,2P}$) deposited in the Galactic bulge as dynamical friction removes massive clusters from the in-situ halo.  The orange symbol-free line depicts the total mass in 2P stars that the in-situ halo contains.  
For all models, $F_{bound}^{VR}=0.4$ and $m_{th}=10^6\,\Ms$. 
}
\label{fig:2PHFS_mup_DF}
\end{center} 
\end{figure} 

The top panel of Fig.~\ref{fig:2PHFS_mup_DF} depicts how dynamical friction raises the field star fraction of the halo, an effect that stems from the decrease of the halo total mass as some of the most massive clusters are transfered to the bulge.  The field star fraction corresponding to the upper limit $\left. F_{2P,HFS}^{12\,Gyr}\right]^+$ (blue line with asterisks, $F_{FHS,Halo}^{12Gyr}=1$) is not modified.  That is, the cluster mass fraction in the halo remains equal to zero since either way -- secular dissolution or dynamical friction -- all clusters are assumed to be destroyed by an age of 12\,Gyr.

The middle panel shows by how much dynamical friction reduces the upper limit $\left. F_{2P,HFS}^{12\,Gyr}\right]^+$ for $\gamma=3.5$ (dark-cyan line with open triangles) and $\gamma=4.5$ (dark-cyan line with filled circles).  As expected, the decrease is more noticeable when the \gc system is initially more centrally-concentrated (i.e.~larger $\gamma$) and for larger $m_{up}$.    

Finally, the bottom panel of Fig.~\ref{fig:2PHFS_mup_DF} presents the mass in stars ($M_{DF}$, purple lines) and the mass in 2P-stars ($M_{DF,2P}$, dark-cyan lines) that are removed from the halo and deposited in the bulge.  The orange line is the total mass in 2P-stars that the in-situ halo contains.  As such, it defines an upper limit to $M_{DF,2P}$.  As $m_{up}$ increases, $M_{DF,2P}$ tends towards $M_{DF}$ for a given $\gamma$, an effect expected as the most massive clusters are made almost exclusively of 2P-stars.  Based on the present model, we expect at most one bulge star out of $10^{3}$ to have formed in massive halo globular clusters.  Most of these stars should be mildly metal-poor (say [Fe/H]$>-1.5$) as a result of the metallicity gradient characterizing in-situ globular clusters \citep[e.g.][]{par00}.  The bulge, however, should contain a larger fraction of 2P stars originating from disk / bulge globular clusters.  The disk / bulge \gc subsystem is more centrally-concentrated than the halo, and its massive members are thus more likely to be removed by dynamical friction.  \citet{lee18,lee19} indeed detected in the outer bulge stars of metallicity $-0.4 \leq {\rm [Fe/H]} \leq 0.3$ that are Na- and Al-enriched.  They propose such stars to originate from disrupted globular clusters.  Dynamical friction must be one of the mechanisms through which the 2P-star content of massive and relatively metal-rich clusters has been deposited in the bulge.

\section{Summary, Conclusions, and Future Work} \label{sec:conc}
We have presented a model mapping the relation between the pristine-star fraction $F_{1P}$ of Galactic \gcs and their present-day mass $m_{prst}$.  The model gives a good account of the region occupied by the observational data in ($m_{prst}$, $F_{1P}$) space, including the respective locations of inner and outer clusters, of single-population clusters, as well as of Magellanic Clouds clusters.  The model also satisfies the observational constraint of a small fraction (few per cent) of 2P-stars in the Galactic halo field. \\

\noindent Model assumptions are as follow: \\  

-1- Forming \gcs start being polluted in \hhb products once their stellar mass has reached a fixed threshold $m_{th}$.  \\

-2- Once the pollution of a cluster begins, the entirety of its \sfing gas is immediately polluted (even the cluster outskirts in case of a centrally-located source of pollution). That is, the formation of 1P-stars stops, the formation of 2P-stars starts.  The mass in 1P-stars of any gas-embedded multiple-populations cluster is thus $m_{th}$.  Hypotheses (1) and (2) yield the relation between the stellar mass of clusters at the end of their formation, $m_{ecl}$, and their pristine-star fraction, $F_{1P}$, namely $F_{1P}(m_{ecl})=m_{th}/m_{ecl}$ if $m_{ecl}>m_{th}$, $F_{1P}=1$ otherwise (Eq.~\ref{eq:f1p-ecl}).  \\     

-3- As \gcs age and lose their stars, 1P- and 2P-stars are equally-likely to be lost.  That is, once the formation of a cluster is over, its 1P-star fraction $F_{1P}$ remains constant in time.  This is not such an extreme assumption since in some dynamically young globular clusters, neither of the two populations (pristine or polluted) is more centrally-concentrated than the other \citep[][their fig.~15]{lei23}.  To obtain the evolutionary paths of clusters in (mass, $F_{1P}$) space therefore boils down to modeling the time evolution of their mass.  \\

-4- The dynamical evolution of clusters after their formation consists of two consecutive phases: response to gas expulsion (i.e. violent relaxation), and long-term secular evolution.  \\

-5- At the end of violent relaxation, clusters retain a fixed fraction $F_{bound}^{VR}$ of their formed stellar mass $m_{ecl}$, regardless of $m_{ecl}$ \citep[see e.g. fig.~1 in][]{par07} and of the tidal field impact \citep[see fig.~4 in][]{shu19}.  \\

-6- The cluster mass at secular-evolution onset is thus $m_{init}=F_{bound}^{VR} m_{ecl}$, which we coin the cluster initial mass.  At this stage, multiple-populations clusters are more massive than $F_{bound}^{VR} m_{th}$.  Using NGC2419, we estimate  $ F_{bound}^{VR} m_{th} = 4 \cdot 10^5\,\Ms$ (see below).  \\

-7- The mass of secularly-evolving clusters decreases linearly with time \citep[Eq.~12 in][]{bm03}, with their dissolution time-scale $t_{diss}$ depending on their initial mass $m_{init}$ and on the strength of the tidal field in which they evolve \citep[see Eq.~10 of][and also our Eq.~\ref{eq:tdiss}]{bm03}.  We quantify the tidal field via an equivalent orbital radius $R_{eq} = (1-e)D_{apo}$, with $D_{apo}$ the apocentric distance and $e$ the eccentricity of the cluster orbit in the Galaxy.  $R_{eq}$ is thus the radius of a circular orbit yielding the same cluster dissolution time-scale as an orbit of apocentric distance $D_{apo}$ and eccentricity $e$.   \\
 
A key advantage of these simple assumptions is to yield a highly-tractable model, from which the relation between \gc present-day mass and pristine-star fraction on the one hand, constraints on the pristine-star fraction and cluster-mass fraction in the Galactic halo on the other hand, can be inferred.  Such a model also provides the foundations upon which more elaborate models can be developed (e.g. different bound fractions for the 1P- and 2P-populations).  \\

The transition from one evolutionary stage to the next is presented in Fig.~\ref{fig:mth}, which also shows 12\,Gyr-old model tracks for equivalent radii $0.5 \leq R_{eq} \leq 8.0$\,kpc.  
Track portions that stretch horizontally to the left (i.e. towards low cluster masses) depict clusters on the edge of final dissolution (Fig.~\ref{fig:tdiss}).  Recall that evolution proceeds at constant $F_{1P}$.  Although the locus of tracks in $({\rm mass},F_{1P})$ space is also influenced by cluster initial concentration and age uncertainties, these effects prove to be minor compared to the tidal field impact (Fig.~\ref{fig:disc}).  

Because model tracks at a given cluster age unfold from the initial track $F_{1P}=F_{bound}^{VR} m_{th}/m_{init}$ (Eq.~\ref{eq:f1p-init}), a critical step is to anchor the latter in (mass, $F_{1P}$) space, equivalently to estimate $F_{bound}^{VR} m_{th}$.  For this purpose, we use NGC2419, a massive and remote -- hence little evolved -- globular cluster.  Assuming that NGC2419 has lost all its stellar-evolutionary mass losses, but none of its stars through secular evolution, we adjust the initial track (orange track in Fig.~\ref{fig:nofred}) such that NGC2419 sets on the track corrected for stellar-evolution mass losses (green track in Fig.~\ref{fig:nofred}). We infer $F_{bound}^{VR} m_{th}=4 \cdot 10^5\,\Ms$ (see top $x$-axis in Fig.~\ref{fig:nofred}).  For $m_{th}=10^6\,\Ms$, a stellar mass threshold allowing for the formation of a Super Massive Star in a dense stellar environment \citep{gie18}, this yields $F_{bound}^{VR}=0.4$, a bound fraction large enough for clusters to survive violent relaxation \citep[e.g.][]{gey01,bk07,shu19}.      

The median equivalent radius $R_{eq}=(1-e)D_{apo}$ of the observational data set is $\simeq 3.3$\,kpc (hence $D_{apo}=8.2$\,kpc for the sample median eccentricity $e \simeq 0.6$).  When using the dissolution time-scale of \citet{bm03}, however,  the model track going through the cloud of data points has $R_{eq}=1$\,kpc (Fig.~\ref{fig:nofred}).  In other words, for the relation $F_{1P}(m_{ecl})$ adopted here, the \citet{bm03} 's cluster dissolution time-scale must be reduced by a factor $f_{red} = 0.3$ so as to push model tracks to lower cluster masses (compare Figs.~\ref{fig:nofred} and \ref{fig:fred}).  This may imply that \gcs form either primordially segregated, or with a top-heavy stellar initial mass function (Fig.~\ref{fig:imf}).  In either case \citep{hag14,hag20}, the cluster life-expectancy is shorter than for a non-primordially segregated cluster with a "canonical" IMF, as considered by \citet{bm03} .  A third possibility could be that the adopted relation $F_{1P}\propto m_{ecl}^{-1}$ is too steep (see 'Future work').

As one moves to small or large equivalent radii, data points become scarce (Fig.~5).  When $R_{eq}>5.0$\,kpc, this growing scarcity mirrors the cluster number density dwindling with Galactocentric distance \citep[see e.g. fig.~2 in][]{mcl99}.  As for the bottom-left part of the diagram ($R_{eq}<2.0$\,kpc), it hosts clusters with a low pristine-star fraction $F_{1P}$, hence an initially high mass (say, $m_{init}>2 \cdot 10^6\,\Ms$).  For such clusters to form, the mass of the Galactic halo must be high enough (size-of-sample effect).  Thereafter, they must also survive dynamical friction, which destroys massive clusters located too close to the Galactic center (thick cyan lines in Fig.~\ref{fig:fred_fric}).  Both factors -- size-of-sample effect and dynamical friction -- therefore explain the scarcity of clusters with 2P-star fraction higher than 80\,\% (see also Sec.~\ref{ssec:2PHFS_mup}).  Should the Milky Way contain a significantly higher number of globular clusters, the distribution of points in (mass, $F_{1P}$) space would be significantly wider.  This would result from the initial cluster mass function and the cluster system radial density profile being populated up to larger masses and galactocentric distances, respectively, which would also allow more massive clusters to evade disruption by dynamical friction.      

Given the greater sensitivity of low-mass clusters to an external tidal field, clusters with a high pristine-star fraction experience a greater range of shifts to the left in ($m_{prst}$, $F_{1P}$) space than high-mass clusters.  
Because of their initial low-mass ($m_{init} < F_{bound}^{VR} m_{th} = 4 \cdot 10^5\,\Ms$), 12\,Gyr-old single-population clusters must be located in the outer halo ($R_{eq} > 4.5$\,kpc;  see top-panel of Fig.~\ref{fig:inout}), which is confirmed by the observations.

Clusters evolving in a stronger tidal field lose a greater fraction of their mass.  As a result, tracks corresponding to short equivalent radii $R_{eq}$ (i.e. short apocentric distances and/or high eccentricities) are located to the left of the diagram, while those corresponding to large equivalent radii stay close to the initial track.  Such a segregation in ($m_{prst}$, $F_{1P}$) space is indeed observed: clusters with $R_{eq}>3.5$\,kpc tend to be located to the right of those with $R_{eq} \leq 3.5$\,kpc (middle panel of Fig.~\ref{fig:inout}).  We stress that our interpretation differs from that of e.g. \citet{zen19}, according to whom outer clusters tend to be {\it above} inner ones.  

Following up on this line of reasoning, clusters evolving in the weak tidal field of dwarf galaxies should remain close to the initial track $F_{1P}(m_{init})$ as well (assuming formation conditions similar to those of the Galaxy).  This is the case of the Magellanic Clouds clusters, which are located closer to the initial track $F_{1P}(m_{init})$ than the vast majority of the Galactic globular clusters, also as a result of their intermediate age \citep[Fig.~\ref{fig:mf1p_mag}; see also fig.~7 in][]{mil20}.  

A key-assumption of the model is that multiple-populations clusters lose their 1P- and 2P-stars with the same likelihood.  A naive expectation may thus be that the stellar field of the Galactic halo contains comparable amounts of 1P- and 2P-stars.  This, however, would contradict the observations that 2P-stars represent at most a few per cent of the halo field stars \citep{car10b,mar11}.  Here, one should keep in mind that 2P-stars form exclusively in clusters with an initial mass $m_{init}>F_{bound}^{VR} m_{th}= 4 \cdot 10^5\,\Ms$ (for the $F_{1P}(m_{ecl})$ relation assumed in Eq.~\ref{eq:f1p-ecl}).  Owing to their high mass, not only are multiple-populations clusters more resilient to secular evolution than single-population clusters, they also represent a limited fraction of the cluster mass spectrum.   We therefore expect secular evolution to shed in the field mostly 1P-stars.  Assuming, for the sake of simplicity, a power-law initial cluster mass function, we have constrained the fraction of 2P-stars in the halo field by considering two extreme scenarios.  In the first one, all clusters dissolve by an age of 12\,Gyr.  This maximizes both the fraction of field stars in the halo ($F_{HFS,Halo}^{12Gyr}=100$\,\%) and the fraction of 2P-stars among halo field stars $F_{2P,HFS}^{12Gyr}$ (blue lines with asterisks in Fig.~\ref{fig:2PHFS_Fb}).  In a second scenario, only single-population clusters secularly dissolve, and the only 2P field stars are those released by violent relaxation.  This minimizes the 2P-star fraction in the halo field (dash-dotted magenta lines in Fig.~\ref{fig:2PHFS_Fb}).  Both scenarios also include an accreted component, modeled as dwarf galaxies made of 1P-field stars \citep{nor17,sal19} and 2P-star clusters.  The in-situ and accreted components are assumed to contribute equally to the total halo stellar mass by an age of 12\,Gyr.  

The model lower and upper limits bracket the observational constraints (horizontal thick black lines in Figs~\ref{fig:2PHFS_Fb} and \ref{fig:2PHFS_mup}) provided that the upper limit on the embedded-cluster mass $m_{up}$ is no larger than $\simeq 7 \cdot 10^6\,\Ms$ (Fig~\ref{fig:2PHFS_mup}).  This condition likely results from the finite Galactic halo mass (size-of-sample effect).  It is also consistent with the scarcity of 2P-star-dominated clusters (i.e. $F_{1P}<<0.2$).  In our model, \gcs therefore do not need to be much more massive at birth than they are today.  This stems from our asssumption that 2P stars form when clusters are still pervaded by their star-forming gas.  Adding the impact of dynamical friction to both scenarios improves how tightly the model lower and upper limits bracket the observations, although its impact is not large (dark-cyan lines in top and middle panels in Fig.~\ref{fig:2PHFS_mup_DF}).  As a result, we expect only a small fraction of the Galactic bulge stars ($<10^{-3}$) to be 2P-stars originating from disrupted massive halo globular clusters (bottom panel of Fig.~\ref{fig:2PHFS_mup_DF}).  We provide all the anlytical equations that the reader needs to derive, for other model input parameters, the fractions of 2P-stars in the halo field and of field stars in the halo. 

Finally, the size-of-sample effect may also explain why clusters in the Magellanic Clouds tend to be richer in 1P-stars than their Galactic counterparts (Fig.~\ref{fig:mf1p_mag}).  The gas mass available to cluster formation being smaller than in the Galaxy, the initial cluster mass function is sampled up to masses lower than in the Galaxy, thereby producing fewer or no 2P-star-dominated clusters.   \\    

{\it Future work: } In this contribution we have assumed a single-index power law for the initial cluster mass function (Eq.~\ref{eq:eclmf}).  We have also assumed that, once given off by whatever source(s), hot hydrogen-burning products pollute immediately the entirety of their forming host cluster (i.e. 1P- and 2P-stars are two consecutive and distinct stellar generations).  Eq.~\ref{eq:f1p-ecl} follows from this latter assumption combined to the assumption of a cluster mass threshold $m_{th}$ for 2P-star formation.  Equations~\ref{eq:f1p-ecl} and \ref{eq:eclmf} have been designed, respectively chosen, for their conceptual simplicity, hence for the possbility of inferring as many analytical results as possible, so as to grasp how the different aspects of the model impact each other.  

However, the true nature of the initial globular cluster mass function has so far remained elusive.  Its present-day Gaussian shape is a shape of equilibrium \citep{ves98}, and the Galactic halo \gc mass function may thus have started as a Gaussian similar to that today, but with a higher normalization \citep[][their fig.~8]{par07}.  Since a Gaussian mass function equates with a double-index power law when clusters are counted per constant linear mass interval \citep[see top panel of fig.~2 in][]{mcl96}, this would imply that low-mass single-population clusters contribute less to the cluster mass budget.  By how much would this affect the 2P-star fraction in the halo field?  

The 2P-field star fraction also depends on whether clusters release their 1P- and 2P-stars equally likely, or not.  In other words, the 2P-field star fraction also depends on whether the 1P- and 2P-populations are similarly distributed inside clusters, or whether one population is more centrally-concentrated than the other.  Recent observations have revealed among dynamically young \gcs the full range of behaviors \citep[fig.~15 in][]{lei23}.  This justifies our present assumption of clusters losing their 1P- and 2P-stars equally likely, while also calling for a generalization of the model with population-dependent bound fractions, especially at violent relaxation, when both populations, if formed segregated, have not had the time to mix yet.  We note that we still lack cluster formation scenarios able to explain all three possible configurations, namely, 1P-population more concentrated than the 2P-population, or vice versa, and 1P- and 2P-populations similarly distributed inside their host cluster.   

Another limitation of the present model is that the protocluster pollution in hot hydrogen-burning products cannot be instantaneous.  If the source of pollution is centrally located, as expected for massive stars, regardless of their mass -- Interacting Massive Binaries \citep{deM09}, Very Massive Stars \citep{vin18,hig23} or Super Massive Stars \citep{den14,gie18} -- one expects 1P-stars to keep forming, at least temporarily, in cluster outer regions, while 2P-stars start forming in inner regions.  Due to this time overlap, 1P- and 2P-stars no longer constitute two fully distinct generations, and the mass in 1P-stars of a cluster $m_{1P}$ does no longer equate with the threshold mass $m_{th}$ for 2P-star formation.  This will necessarily modify the $F_{1P}(m_{ecl})$ relation adopted here (Eq.~\ref{eq:f1p-ecl}), with consequences for the factor $f_{red}$ by which the cluster dissolution time-scale of \citet{bm03} has to be reduced (Eq.~\ref{eq:tdiss}).  

Mapping the observed distributions of points in additional parameter spaces \citep[e.g. $F_{1P}(m_{1P})$, $F_{1P}(m_{2P})$, and $F_{1P}(m_{init})$; see top-right and bottom panels of fig.~7 in][]{mil20} is also part of the tasks that we will pursue in forthcoming papers.    

\acknowledgments
GP acknowledges funding by the Deutsche Forschungsgemeinschaft (DFG, German Research Foundation) -- Project-IDs 515414180; 138713538, SFB 881 ("The Milky Way System", subproject B05, PI: Dr.~A.~Pasquali). This research was also supported by the M\"unich Institute for Astro-, Particle and BioPhysics (MIAPbP), which is funded by the Deutsche Forschungsgemeinschaft under Germany   's Excellence Strategy -- EXC-2094 -- 390783311.  GP is grateful to H.~Baumgardt and M.~Hilker for having provided her with their GC mass estimates in tabular format at the MIAPbP Workshop "{\it Star-Forming Clumps and Clustered Starbursts across Cosmic Time}".  She is also grateful to an anonymous referee for a careful and constructive report.  This research has made use of NASA 's Astrophysics Data System.



\appendix
\section{About the assumed constancy of the bound fraction $F_{bound}^{VR}$}
\label{sec:apA}

As mentioned in Sec.~\ref{ssec:vr}, the bound fraction of stars retained by clusters at the end of violent relaxation, $F_{bound}^{VR}$, depends on several cluster properties at gas expulsion onset, including their star formation efficiency, gas-expulsion time-scale, and the tidal field impact.  Therefore, in this Appendix, we discuss how sensible our assumption of a bound fraction common to all clusters is, even with expected cluster-to-cluster variations of the star formation efficiency, gas expulsion time-scale, and tidal field impact.  

The bound fraction stretches from $0$ (complete cluster dissolution following gas expulsion) to $1$ (cluster left unaffected by gas expulsion).  However, as soon as the \sfe becomes high enough (say $SFE>0.5$) and/or gas expulsion becomes slow enough (say, longer than two embedded-cluster crossing times $\tau_{cross}$), the bound fraction remains trapped in a fairly narrow range, i.e. $0.4 \lesssim F_{bound}^{VR} \lesssim 1.0$  \citep[see figs~1-2 in][]{bk07}.  Such conditions can be achieved in dense and compact cluster-forming clumps, whose short crossing-time and deep potential well hinder the fast removal of the star-forming gas, thereby allowing high star formation efficiencies to be achieved.  

The tidal field impact is usually quoted as the ratio between the half-mass radius and tidal radius of the embedded cluster, $r_{hm}/r_{tidal}$ \citep[i.e. the cluster is more vulnerable to the external tidal field for larger $r_{hm}/r_{tidal}$; see e.g.][]{bk07}. 
A tidal field impact common to all clusters can be set if the density $\rho_{ecl}$ of forming clusters scales with the density of their environment $\rho_{gal}$ since $r_{hm}/r_{tidal} \propto (\rho_{gal}/\rho_{ecl})^{1/3}$.  Moreover, when star clusters form with a constant \sfe {\it per free-fall time}, the \sfe is locally higher in the cluster dense central regions than in their diffuse outskirts \citep[see fig.~10 in ][]{par13}.  Such a centrally-peaked profile of the local \sfe inside clusters results in a bound fraction $F_{bound}^{VR}$ that depends only weakly on the tidal field impact $r_{hm}/r_{tidal}$ \citep[see figs~4-5 in][]{shu19}.      

We therefore stress that a bound fraction $F_{bound}^{VR}$ common to all clusters does not necessarily imply a common star formation efficiency, gas-expulsion time-scale and tidal field impact for all clusters.  The formation conditions of \gcs may well conspire for their bound fraction not to vary widely among clusters, for instance because their \sfe is high enough, gas expulsion is slow enough, and either the tidal field impact is negligible (as a result of a high density contrast between clusters and their environment), or its variations hardly affect $F_{bound}^{VR}$ (as a result of cluster formation proceeding with a constant \sfe per free-fall time).  

For the sake of simplicity, we shall therefore assume a fixed bound fraction, and we adopt $F_{bound}^{VR}=0.40$.

\section{Haghi et al 's (2020) assumptions lead to $F_{bound}^{VR} \simeq 1$ for massive clusters}
\label{sec:apB}
In Appendix~B, we demonstrate that, following \citet{hag20} 's assumptions, massive clusters lose a minor fraction of their stars during violent relaxation (i.e. $F_{bound}^{VR} \lesssim 1$ and $m_{init} \lesssim m_{ecl}$).  The reason for this is two-fold: adiabatic residual \sfing gas expulsion and weak tidal-field impact.  

Let us first demonstrate the adiabaticity of gas expulsion, for which we need the gas expulsion time-scale in units of the gas-embedded cluster crossing-time.  By combining the gas-expulsion time-scale $t_{decay}$ of \citet[][their Eq~4]{hag20}, their relation of the cluster half-mass radius in dependence of $m_{ecl}$ (their Eq.~1), their assumed \sfe $SFE=0.33$, and the expression of the gas-embedded cluster crossing-time $t_{cross}$ for a \citet{plu1911} model \citep[Eq.~6 in][]{bk07}, we obtain $t_{decay}/t_{cross} = 0.009m_{ecl}^{0.43}$.  This yields $t_{decay}/t_{cross} \simeq 3.5$ for $m_{ecl}=10^6\,\Ms$ and, therefore, following \citet{hag20} 's assumptions, massive clusters expel their residual \sfing gas adiabatically.  The cluster expansion factor during violent relaxation is thus $\simeq SFE^{-1}$ \citep[][]{mat83}, which is here $SFE^{-1} \simeq 3$.

In turn, this limited expansion allows massive clusters to remain well-shielded against the Galactic tidal field until the end of their violent relaxation.  Table~A1 of \citet[][]{hag20} shows that the ratio between their half-mass radius and tidal radius is $<0.01$ at the gas-embedded stage.  This result can be recovered by combining Eqs~1 and 6 of \citet{hag20}: we find $(r_{hm}/r_{tidal})_{ecl}=0.067m_{ecl}^{-0.2}$, which is less than 0.01 in the high-mass regime.   
At the end of violent relaxation, clusters are still well inside their tidal radius since the ratio between their half-mass- and tidal radius is then $\lesssim 0.03$.  As a result, $F_{bound}^{VR} \lesssim  1$.   

This outcome is supported by the equal-mass particles $N$-body simulations of \citet{bk07}, in which they find a bound fraction $F_{bound}^{VR} \simeq 0.90$ when $SFE=0.33$, $\tau_{decay}/\tau_{cross} > 1$ and $(r_{hm}/r_{tidal})_{ecl}=0.01$ (see the top-left panel of their Fig.~2).  Additionally, Fig.~4 of \citet{hag20} shows that, for a Kroupa IMF, the cluster dissolution time-scales $t_{diss}^{BM03}(m_{init})$ of \citet{bm03} and $t_{diss}^{HS20}(m_{ecl})$ of \citet{hag20} are very close to each other in the high-mass regime, while, in the low-mass regime, the relation $t_{diss}^{BM03}(m_{init})$ gets shifted to lower mass with respect to the relation $t_{diss}^{HS20}(m_{ecl})$, thereby showing that during violent relaxation low-mass clusters lose more stars than high-mass ones.  

We conclude therefore, that, following \citet{hag20} 's assumptions, high-mass clusters -- those forming $2P$ stars -- satisfy $F_{bound}^{VR} \lesssim 1.0$ and $m_{init} \lesssim m_{ecl}$.

\end{document}